# Zero photoemission from icosahedral superatoms

Yuta Suzuki[1], Mamoru Tanaka[1], Yoshi-ichi Suzuki[2], Akira Terasaki[1,3]*, Takuya Horio[1]*

[1]Department of Chemistry, Faculty of Science, Kyushu University; 744 Motooka, Nishi-ku, Fukuoka 819-0395, Japan
[2]School of Medical Technology, Health Sciences University of Hokkaido; 1757 Kanazawa, Tobetsucho, Ishikari, Hokkaido 061-0293, Japan
[3]Quantum and Spacetime Research Institute (QuaSR), Kyushu University; 744 Motooka, Nishi-Ku, Fukuoka 819-0395, Japan

*Corresponding authors. Email: terasaki@chem.kyushu-univ.jp,
horio@chem.kyushu-univ.jp

## Abstract

Photoemission is expected to occur when the photon energy exceeds the binding energy of an electron, with the photoemission yield generally remaining non-zero above this threshold. Yet, it can be suppressed at specific photon energies, a phenomenon known as the Cooper minimum. Although complete suppression represents an extreme limit of this phenomenon, such zero photoemission has previously been reported only for atoms with an outermost s-type orbital and has never been observed in non-atomic systems. Here, we demonstrate the deliberate realization of zero photoemission from $Ag_{12}W^-$, an artificial icosahedral superatom, by observing a vanishing photoemission cross section 0.85 eV above the detachment threshold and reproducing this behavior in theoretical simulations. These simulations predict that zero photoemission can be common in nearly spherical nanoclusters, a prediction we experimentally confirm using $Ag_{12}Mo^-$, an isoelectronic counterpart of $Ag_{12}W^-$. Because superatoms offer systematic control over both size and elemental composition using the periodic table, our findings establish a route toward active control of photoemission in finite-size quantum systems and stimulate further investigations of underexplored physical phenomena such as nondipole effects.

Photoemission, which is expected whenever the photon energy exceeds the electron binding energy, is among the most fundamental processes in nature. In general, the photoemission yield remains non-zero above the binding-energy threshold. Yet, more than six decades ago, Cooper predicted that the photoemission yield can be suppressed at certain photon energies or photoelectron kinetic energies (PKEs)[1] (Fig. 1A). This phenomenon is known as the Cooper minimum (CM) that originates from a sign reversal of the PKE-dependent, radial component of the transition dipole matrix element (TDME). To date, CMs have been observed as minima in energy-dependent photoionization cross sections[2,3] or high-harmonic spectra of atoms[4,5]. In non-atomic systems, however, only a few observations have been reported, involving lone-pair orbitals[6] and Rydberg states of molecules[7], or strongly localized transition-metal d orbitals at solid surfaces[8]. All these systems retain pronounced atomic-like character.

In these non-atomic systems, the photoemission yield reaches a minimum but does not completely vanish. This is because photoelectron emission proceeds through multiple partial waves, making it highly unlikely for all corresponding TDMEs to vanish simultaneously at the same PKE. This suggests that, if photoelectron emission can be restricted to a single partial wave, a CM need not merely be a minimum in the photoemission yield: the photoemission amplitude can, in principle, vanish completely. Such complete suppression of photoemission (i.e., zero photoemission) can occur for an atomic s orbital, for which the dipole selection rules restrict the outgoing electron to a p-wave. Nevertheless, this phenomenon is extremely rare even in atomic systems[2,9–11] and has never been reported in non-atomic systems. The PKE at which zero photoemission occurs is exceedingly sensitive to the radial overlap between bound and continuum wavefunctions, making this phenomenon a stringent probe of their shapes and nodal structures. Zero photoemission can also provide a key platform for investigating photoemission time delays[12,13] and nondipole effects[14–16]. Here, we demonstrate the deliberate realization of zero photoemission by engineering an almost perfectly spherical bound-state wavefunction in an artificial icosahedral nanocluster, or superatom, through appropriate control of the number and elemental composition of its constituent atoms.

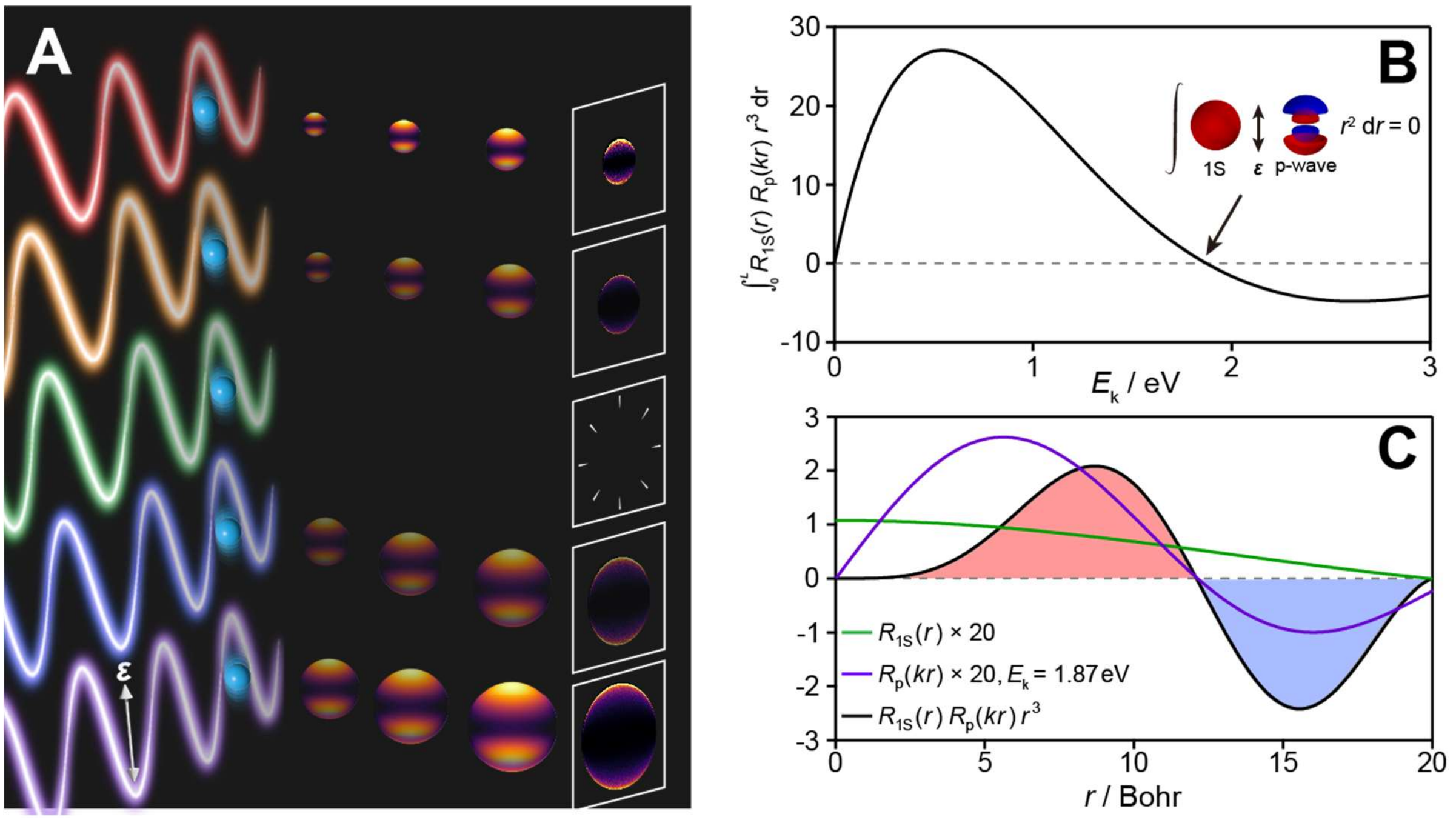

**Fig. 1. Conceptual illustration of zero photoemission via the Cooper-minimum mechanism.** (A) Schematic of photoemission from atoms as a result of changing laser wavelengths. In this example, photoelectrons are emitted along the laser polarization vector, $\varepsilon$, and projected onto an imaging detector plane, in which the photoemission yield vanishes at a specific laser wavelength. (B) Radial component of transition dipole matrix elements (TDMEs) from a 1S superatomic orbital, calculated as a function of the photoelectron kinetic energy (refer to the text), highlighting the zero crossing at $E_k = 1.87$ eV of the TDME. (C) Radial wavefunctions of the 1S orbital (green) and the outgoing p-wave at $E_k = 1.87$ eV ($k = 0.371$ Bohr$^{-1}$; purple), along with the TDME integrand (black). The red and blue hatched regions represent the positive and negative integral, respectively.

Superatoms are generally defined as aggregates of atoms that mimic single atoms. Metal superatoms are known to exhibit atomic-like electronic shell structures, which can be described by the spherical jellium model[17]. The eigenstates of this model, often referred to as superatomic orbitals (SAOs), are labeled 1S, 1P, 1D, 2S,…, using the principal quantum number and the orbital angular momentum quantum number, in analogy with atomic orbitals (1s, 2s, 2p, 3s, 3p, 3d, …).

Figure 1B shows a numerically calculated TDME for photoemission from a 1S SAO, where the radius of the spherical potential confining the electron was arbitrarily set to 20 Bohr to represent the situation shown in Fig. 1A, and the continuum-state wavefunction was approximated by a plane wave. Under perfect spherical symmetry, photoemission proceeds exclusively through the p-wave channel within the dipole approximation. The calculated TDME exhibits a zero crossing at a PKE of 1.87 eV, resulting in complete suppression of the photoemission cross section. As shown in Fig. 1C, this results from the exact cancellation of the positive (red) and negative (blue) contributions to the TDME, representing zero photoemission, an extreme case of a CM. Our simple numerical simulation therefore suggests that zero photoemission can be realized in highly symmetric nanoclusters.

The key challenge is to synthesize an ideal superatom that closely approximates spherical symmetry and possesses an S-type SAO as its valence orbital. Promising candidates include metal nanoclusters with icosahedral ($I_h$) symmetry—the highest symmetry among finite point groups—such as Mackay icosahedra[18] consisting of 13, 55, 147, ... atoms. Inspired by pioneering studies of doped metal clusters[19,20], we recently revealed that cluster anions of $Ag_{12}M^-$ (M = V, Nb, Ta) all adopt $I_h$ structures with the Group 5 dopant atom encapsulated at the center and closed electronic shells of $1S^2 1P^6 1D^{10}$ (Ref.[21]). Accordingly, the ground electronic states of $Ag_{12}M^-$ (M = V, Nb, Ta) are $S_0(^1A_g)$. Based on this finding, we expected that doping $Ag_{12}^-$ with a Group 6 atom (e.g., Cr, Mo, W) would yield an $I_h$ superatom with a $1S^2 1P^6 1D^{10} 2S^1$ configuration ($^2A_g$). This provides an ideal system in which the cluster symmetry can be maintained without substantial structural change such as Jahn-Teller distortion upon the photodetachment of $^1S_0(^1A_g) + e^- \leftarrow {}^2D_0(^2A_g)$, thereby mimicking photoemission from an atomic s orbital.

We first performed geometry optimization of $Ag_{12}W^-$ using density functional theory (DFT) calculations and found that the $I_h$ form is indeed the most stable, with the W atom encapsulated in the center of the icosahedral $Ag_{12}$ cage (Fig. 2A); the computational results are summarized in Fig. S3 of the Supplementary Information. We confirmed that no structural isomers are present within ~0.8 eV relative to the $I_h$ form; therefore, their coexistence can be ignored in our setup. Fig. 2A highlights that the singly occupied molecular orbital (SOMO) of $Ag_{12}W^-$ is clearly of superatomic S character with no nodes in angular directions. We then produced $Ag_{12}W^-$ in the gas phase and obtained its photoelectron image at a photon energy of 1.66 eV (refer to Fig. S1 in Supplementary Information for details of the experimental methods). The image (Fig. 2B) shows a large positive anisotropy along the laser polarization, $\varepsilon$, which is a characteristic behavior for photoemission from s(S) orbitals.

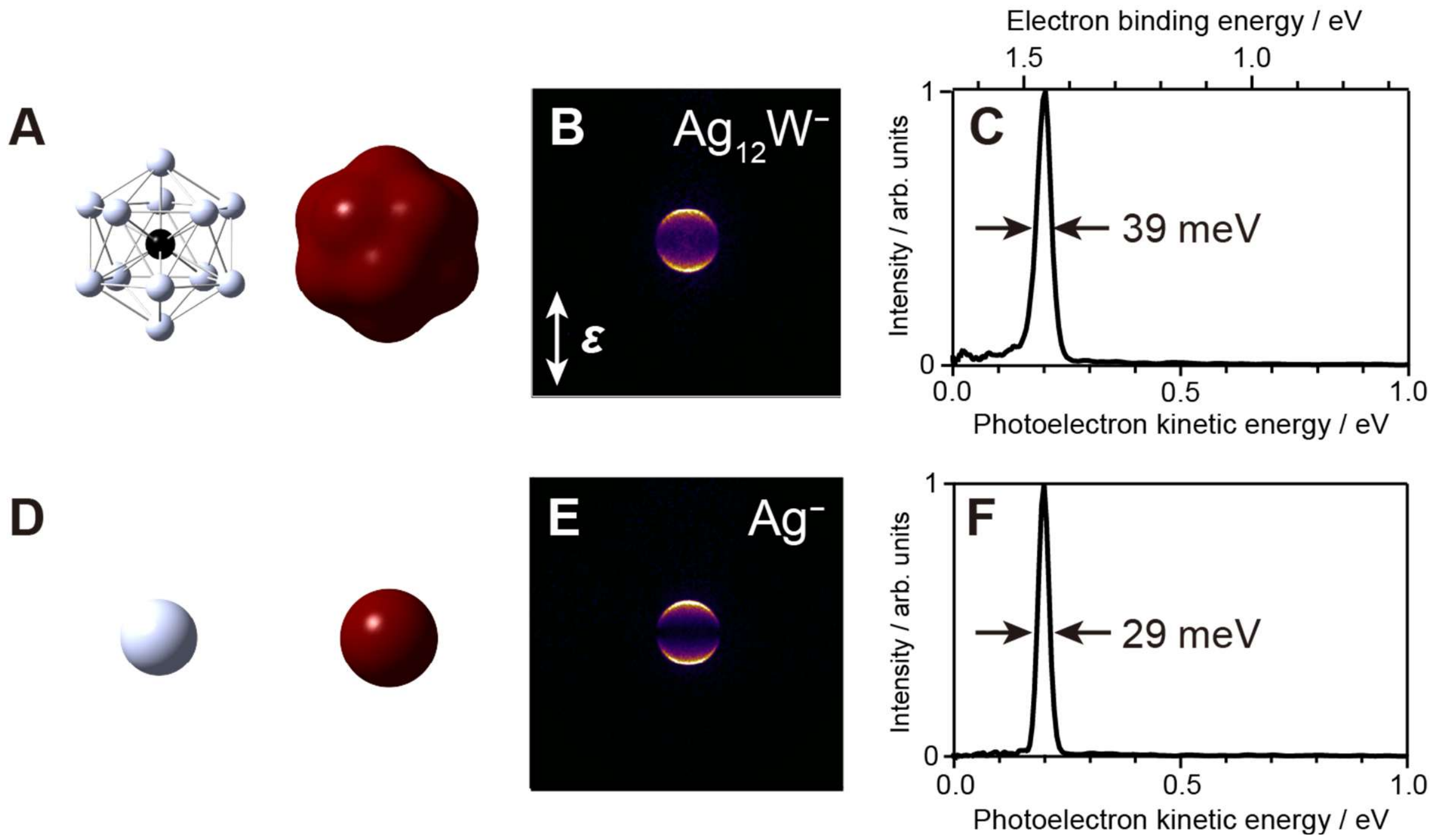

**Fig. 2. Laboratory-frame photoelectron angular distribution and photoelectron spectrum reflect the atom-like character of $Ag_{12}W^-$ upon photodetachment from SOMO.** (A) Optimized geometry of $Ag_{12}W^-$ and the isosurface plot of its SOMO (S-like orbital) obtained at the PBEPBE/LANL2DZ level of theory (isovalue = 0.010 a.u.). Ag and W atoms are displayed as white and black spheres, respectively. (B) Photoelectron image of $Ag_{12}W^-$ obtained at a photon energy of 1.66 eV, and (C) the corresponding photoelectron spectrum. (D) $Ag^-$ and an isosurface plot of its 5s orbital. (E) Photoelectron image of $Ag^-$ obtained at a photon energy of 1.50 eV and (F) the corresponding photoelectron spectrum.

The photoelectron spectrum of $Ag_{12}W^{-}$ (Fig. 2C) yields a vertical detachment energy (VDE) of 1.46 eV. The calculated VDEs are summarized in Table S1 in the Supplementary Information, and the value obtained at the PBEPBE/LANL2DZ level of theory (1.46 eV) is in excellent agreement with the experimental value. The spectrum is dominated by a single peak with a full-width at half-maximum of 39 meV, which is close to our instrumental energy resolution (Fig. 2F). This result indicates that structural relaxation upon the photodetachment of the $^{1}S_{0}(^{1}A_{g}) + e^{-} \leftarrow {}^{2}D_{0}(^{2}A_{g})$ transition is negligible. In fact, our DFT calculation indicates that neutral $Ag_{12}W$ also prefers the $I_h$ form (Fig. S4), fully consistent with our intuitive prediction and the previous computational results[22,23]. A comparison of Fig. 2B with a photoelectron image (Fig. 2E) obtained for the silver atomic anion, $Ag^{-}$, emitting photoelectrons from its 5s orbital shows a marked similarity between $Ag_{12}W^{-}$ and $Ag^{-}$. That is, $Ag_{12}W^{-}$ possesses an S-type SAO and also retains a rigid framework upon photodetachment, making it an ideal superatomic system for investigating zero photoemission.

We recorded 55 images (i.e., two-dimensional projections of photoelectron distributions from $Ag_{12}W^{-}$) by varying the photon energy from 1.46 to 4.00 eV; the results are summarized in Fig. 3. The total photoelectron counts for each image were normalized by the total number of photons in the incident laser beam and the anion beam intensity; they therefore reflect the relative photodetachment yield (Fig. 4A). Notably, different dynamic ranges of photoelectron counts are used for the first, second, third, and subsequent rows to represent the large variation in the photoelectron yield over the measured photon energy range. We highlight that our high-repetition-rate photoelectron imaging setup[24–26] enables fast image acquisition from such dilute samples with an extremely low number density ($\sim 10^{3}$ cm$^{-3}$). With increasing photon energy from the threshold, the photoelectron signals expand and exhibit a ring-like signature with pronounced anisotropy along the laser polarization axis, $\varepsilon$. However, from 2.11 to 2.48 eV, the photoelectron ring is almost invisible, suggesting zero photoemission. When the photon energy is greater than 2.53 eV, an anisotropic photodetachment signal reappears, and additional features emerge in the range 2.89–3.18 eV, where the anisotropy of the outermost ring becomes much less pronounced, before recovering to the strong positive anisotropy when the photon energy exceeds 3.35 eV. The slice images corresponding to Fig. 3 (**Extended Data Fig. 1**) should be helpful in understanding the drastic anisotropy changes of the outermost ring. For a quantitative comparison of the photoelectron intensities, the corresponding photoelectron spectra are provided in **Extended Data Fig. 2**. The integrated intensity of each spectrum provides the relative photodetachment yield.

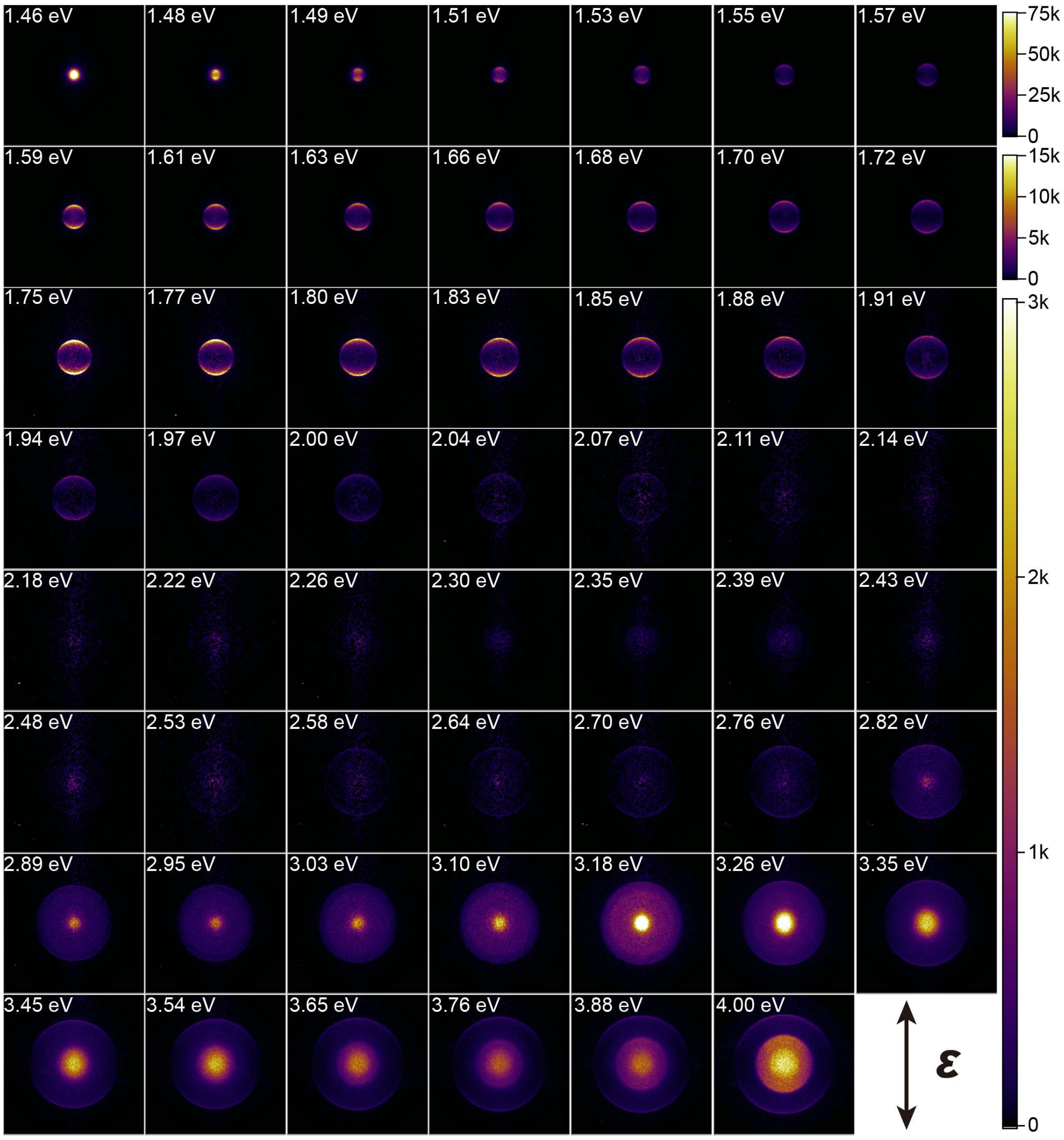

**Fig. 3. Photoelectron images (2D projections) of $Ag_{12}W^-$ obtained at photon energies from 1.46 to 4.00 eV**. The polarization vector, $\varepsilon$, of the photodetachment laser is indicated by a double arrow in the bottom-right corner. The color scales of images in the first, second, third, and subsequent rows are shown on the right. The corresponding slice images are provided in **Extended Data Fig. 1**.

The red solid circles in Fig. 4A show the photoelectron yield spectrum of $Ag_{12}W^-$, as obtained by integrating the total photoelectron signals from each image in Fig. 3. In addition, using a different experimental setup[27], we acquired the photodetachment action spectrum in the photon energy range from 1.46 to 5.51 eV. We monitored the depletion of the number of trapped $Ag_{12}W^-$ anions upon photoirradiation in a linear ion trap (refer to Fig. S2 in Supplementary Information for the experimental details). The result is shown as light-blue solid circles in Fig. 4A. The two spectra, which were obtained by independent experimental methods, are in excellent agreement in the photon energy range from 1.46 to 4.0 eV, which is the highest photon energy available in the photoelectron imaging setup. Above 1.46 eV, the photodetachment cross section decreases gradually with increasing photon energy and eventually becomes almost zero in the photon energy range 2.0–2.5 eV (inset of Fig. 4A). As the photon energy further increases from 2.5 eV, the photodetachment cross section recovers, followed by a small peak that appears at 3.2 eV and a huge, broad band that appears at 4.3 eV.

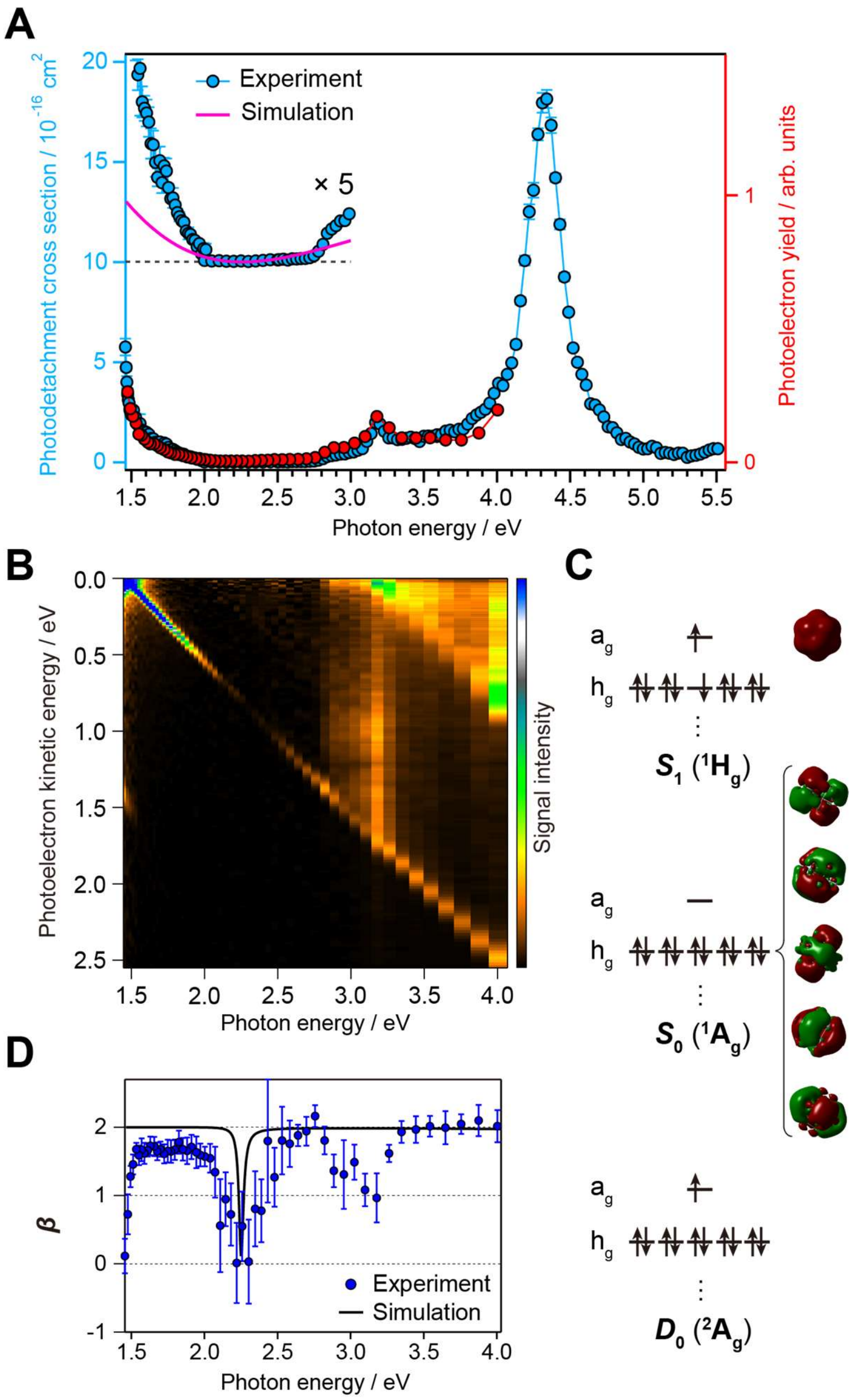

**Fig. 4. Characterization of the zero photoemission from $Ag_{12}W^{-}$ by photoelectron imaging and photodetachment action spectroscopy.** (A) Experimental photodetachment action spectrum (light-blue circles) and photoelectron yield spectrum (red circles) as a function of the photon energy. The inset shows a 5× enlarged view of the experimental photodetachment cross section (light-blue circles) and simulated photodetachment cross sections by ezDyson for $S_0 + e^- \leftarrow D_0$ channel (pink solid line). (B) Two-dimensional map of photoelectron spectra versus photon energy; the integrated photoelectron spectrum at each photon energy is proportional to the red circles shown in (A). (C) Leading electronic configuration of $S_1(^1H_g)$, $S_0(^1A_g)$, and $D_0(^2A_g)$ states for neutral $Ag_{12}W$ and anionic $Ag_{12}W^{-}$. (D) Photoelectron anisotropy parameters, $\beta$, for the $S_0 + e^- \leftarrow D_0$ transition of $Ag_{12}W^{-}$ as a function of photon energy (circles: experiment; solid line: ezDyson simulation).

Figure 4B presents a two-dimensional (2D) map of photoelectron spectra extracted from the 55 photoelectron images shown in Fig. 3; the horizontal and vertical scales are the incident photon energy and PKE, respectively. The diagonal component that emerges at a photon energy of 1.46 eV and extends toward the lower right of the map is assigned to the $S_0(^1A_g) + e^- \leftarrow D_0(^2A_g)$ transition (i.e., one-photon detachment from the S-like ($a_g$) orbital; refer to Fig. 4C for the corresponding electronic configurations). Its relative intensity is clearly vanishing at approximately 2.3 eV in the photon energy (i.e., ~0.85 eV in PKE), manifesting the emergence of zero photoemission. Intense signals are evident in the upper-right region of the map over the photon energy range 3.2–4.0 eV (i.e., 0–0.8 eV in PKE); these signals are assigned to the $S_1(^1H_g) + e^- \leftarrow D_0(^2A_g)$ transition, corresponding to photodetachment from the fivefold-degenerate, D-like ($h_g$) orbital (Fig. 4C). The blue circles in Fig. 4D represent the anisotropy parameter, $\beta$, obtained for the $S_0(^1A_g) + e^- \leftarrow D_0(^2A_g)$ transition. From the threshold to 1.6 eV, the $\beta$ value increases steeply from 0 to approximately 1.7, then remains nearly constant until the photon energy reaches 2.0 eV. The experimental $\beta$ values exhibit a dramatic decrease to nearly zero between 2.0 and 2.5 eV, which coincides with the vanishing photodetachment yield observed in Figs. 4A and 4B.

We conducted theoretical simulations for these experimental results using the ezDyson program[28,29]. Although Dyson orbitals are required for rigorously evaluating TDME, their calculations are computationally too demanding for the present system. Hence, we substituted a Kohn–Sham orbital (Fig. 2A) for a Dyson orbital. For a photoelectron wavefunction, we assumed a Coulomb wave with the effective charge, $Z_{eff}$, as a variable parameter because a plane wave (i.e., $Z_{eff} = 0$), which is commonly used for photodetachment from anions, failed to reproduce the experimental results (see Supplementary Information for further details of the simulation). The pink curve in Fig. 4A and the black curve in Fig. 4D represent the simulated photodetachment cross section and $\beta$ values, respectively, for the $S_0(^1A_g) + e^- \leftarrow D_0(^2A_g)$ transition, as obtained with $Z_{eff}$ = +0.37$e$. The simulated cross section gradually decreases from the threshold to 2.0 eV, becomes almost zero in the photon energy range of 2.0–2.5 eV, and recovers as the photon energy increases; notably, the simulated curve exhibits a non-zero value at the threshold because we used a Coulomb wave. The theoretical $\beta$ values fully reproduce the drastic decrease at ~2.3 eV that coincides with the vanishing photodetachment yield. Although the present approach is phenomenological, the simulations well reproduce both of the experimental results consistently.

We now examine the radial component of the TDME for the $S_0(^1A_g) + e^- \leftarrow D_0(^2A_g)$ process in detail to elucidate the origin of the observed zero photoemission. The result is presented in Fig. 5. The green curve shows the SOMO of $Ag_{12}W^-$ averaged over the full solid angle at each radius, $r$ (i.e., the radial part of the wavefunction). A partial view of the cluster's $I_h$ framework with an isosurface plot of the SOMO (isovalue = 0.010 a.u.) is also presented in the background of Fig. 5, showing that a W atom, located at the center of the cluster (i.e., at $r$ = 0 Bohr), is surrounded by Ag atoms. The purple curve shows a Coulomb wave with $Z_{eff}$ = +0.37$e$ to describe the photoelectron wavefunction with a PKE of 0.85 eV, at which the photoemission yield is vanishing. Multiplying the green and purple curves and further multiplying by the factor $r^3$ yields the black curve in Fig. 5, representing the radial part of the TDME. The positive and negative contributions of the integrand exactly cancel, yielding a zero integral. Thus, this numerical examination confirms that the vanishing photoemission cross section originates from zero photoemission, an extreme case of a CM.

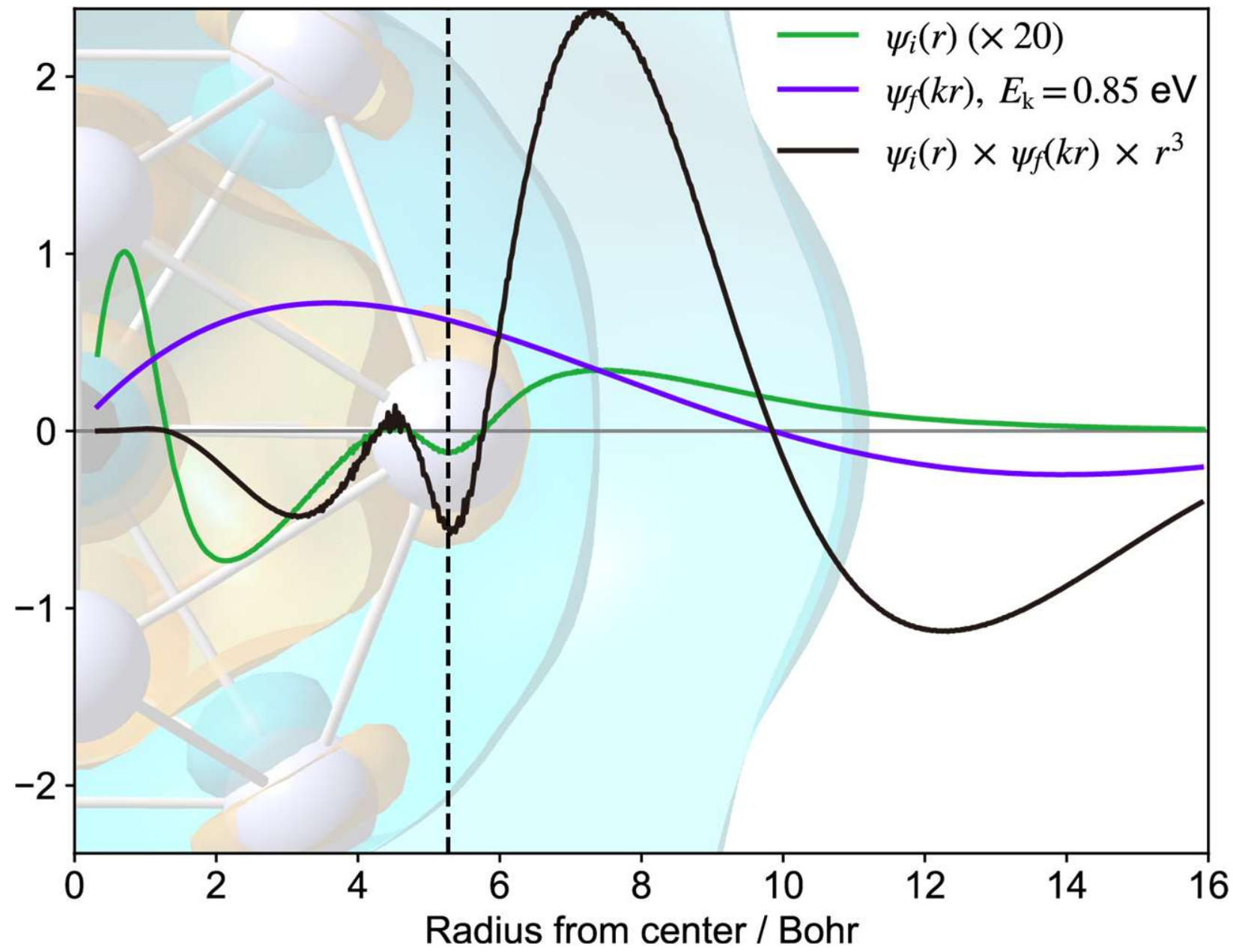

**Fig. 5. Calculation of the radial component of the transition dipole moment accounting for the zero photoemission.** Calculated radial wavefunction (20× enlarged view) of the SOMO of $Ag_{12}W^-$ (green), Coulomb wavefunction with +0.37*e* charge for a photoelectron wavefunction with 0.85 eV photoelectron kinetic energy (purple), and integrand of the transition dipole matrix element (black). The isosurface plot of the SOMO (isovalue = 0.010 a.u.) is shown in the background for correspondence to the radial wavefunction of the SOMO, along with the W–Ag bond length (black dashed line, 5.27 Bohr). Notably, the first node discernible in the radial wavefunction (approximately 1.2 Bohr) is attributable to the 6s atomic orbital of a W atom.

The above analysis, combined with the concept of SAOs (i.e., 1S, 1P, 1D, 2S,…), further predicts that zero photoemission can be induced by engineering nearly spherical bound-state wavefunctions in superatoms through careful control of the numbers and elemental compositions of constituent atoms. We experimentally confirm this prediction by producing $Ag_{12}Mo^-$, an isoelectronic counterpart of $Ag_{12}W^-$; the outermost electronic shell of $Ag_{12}Mo^-$ is therefore expected to be 2S as well. As shown in **Extended Data Fig. 3**, the photoelectron spectrum of $Ag_{12}Mo^-$ also exhibits a single, sharp peak. This spectral feature indicates that the Mo-doped cluster anion adopts $I_h$ symmetry, as confirmed by DFT geometry optimization at the PBEPBE/LANL2DZ level of theory; the calculated result is summarized in Fig. S10 of the Supplementary Information. **Extended Data Figs. 4A and 4B** show the photoelectron yield spectrum and the 2D map of photoelectron spectra obtained for $Ag_{12}Mo^{-}$, respectively. Zero photoemission is clearly observed at a photon energy of approximately 2.1 eV, corresponding to a PKE of ~0.87 eV.

Before concluding, we revisit Fig. 4D, which shows the experimental and computational $\beta$ values obtained for the $S_0(^1A_g) + e^- \leftarrow D_0(^2A_g)$ photodetachment process of $Ag_{12}W^{-}$. First, we note that the simulation curve deviates from the experimental data over the photon energy ranges 1.46–2.0 and 2.7–3.3 eV, likely because of autodetachment from resonance states that are not included in the ezDyson simulation (see Fig. S6 in Supplementary Information). Most importantly, the atomic limit value of $\beta$ = 2 is obtained at ~2.7 eV and throughout the 3.3–4.0 eV range, demonstrating the fully atom-like character of the present superatom. This contrasts with photodetachment from an S-like orbital of $Na_{19}^{-}$, for which the $\beta$ values do not reach the atomic limit ($\beta \leq 1.7$)[30] because of its low structural symmetry ($C_s$)[31,32]. The atomic-limit value in the present study confirms that, under the assumption of full spherical symmetry, photoelectron emission occurs exclusively through a p-wave with orbital angular momentum quantum number $l$ = 1. By contrast, as shown in Fig. S5C in Supplementary Information, the p-wave contribution vanishes at the photon energy corresponding to zero photoemission, whereas the contributions from the h partial wave ($l$ = 5) become dominant, accounting for the pronounced deviation of the $\beta$ value from 2. The vanishing p-wave contribution thus reveals the extremely weak contribution of the higher-order partial wave, which would otherwise be masked at low PKEs by the Wigner threshold law[33]. Notably, the h partial wave arises from the mixing of the S and I orbital angular momentum components in the SOMO under $I_h$ symmetry (see Table S2 in the Supplementary Information).

## Conclusion

We have demonstrated the deliberate realization of zero photoemission, an extreme case of a CM, by creating a superatomic S orbital in $Ag_{12}W^{-}$. Notably, the angular anisotropy shows the atomic limit ($\beta = 2$) for the photoelectrons emitted from this superatomic S orbital. These fully-atomic behaviors are expected to be common to nearly-spherical $I_h$ nanoclusters. Indeed, we experimentally confirm this prediction by producing $Ag_{12}Mo^{-}$ and obtaining its energy-dependent photodetachment cross section. The vanishing photoemission identified in this study originates from an occupied orbital delocalized over the entire metal cluster and is observed through direct photodetachment without involving an intermediate excited state. This behavior contrasts with previously reported non-atomic cases associated with localized atomic orbitals or Rydberg-state-mediated processes. Because the composition and size of superatoms can be systematically tuned by varying the host element, guest element, and the number of constituent atoms, our findings establish a route toward active control of their photoemission processes and further stimulate investigations of underexplored physical phenomena such as nondipole effects.

## Methods

<u>Photoelectron imaging and photoelectron yield spectroscopy</u>

The whole experimental setup, including a photodetachment light source, has been described in detail in our recent reports[24–26]. A schematic illustration of the experimental apparatus is shown in Fig. S1 in the Supplementary Information. An assembly of a tungsten disk placed underneath a silver disk with two through holes (6.5 mm in diameter) was sputtered in a magnetron-sputtering cluster ion source using argon cations. Both the disks were 99.99% in purity, 50.8 mm in diameter, and 3 mm in thickness. Sputtered tungsten and silver atoms were cooled by helium buffer gas at ~120 K to facilitate clustering. The resulting charged clusters were transported via an octupole ion guide to a quadrupole ion deflector, which selectively transmitted anionic species to a quadrupole mass filter (MAX-4000, Extrel CMS, LLC)[34]. A continuous, mass-selected $Ag_{12}W^{-}$ anion beam was delivered from the exit lens of the mass filter to a potential switch using octupole ion guides. At the potential switch, the continuous beam of $Ag_{12}W^{-}$ was chopped into 100-kHz anion bunches by +0.4-kV pulses synchronized with photodetachment laser pulses as described below.

The bunched $Ag_{12}W^{-}$ anions entered a photoelectron velocity-map imaging (VMI) section holding an assembly of electrodes set at low electric fields (~10 V/cm)[35] to extract photoelectrons. The cluster anions were irradiated by photodetachment laser pulses with the wavelengths ranging from 310 to 852 nm (1.46–4.00 eV). The laser pulses were generated by a combination of an Yb:KGW regenerative amplifier (PHAROS, 10 W, 1030 nm, 210 fs, 100 kHz, Light Conversion), an optical parametric amplifier (OPA) (ORPHEUS, Light Conversion), and an additional beta-barium borate ($\beta$-BBO) crystal for frequency-doubling the OPA outputs. The average power of the photodetachment laser pulses was 50–350 mW depending on the wavelength, which were monitored during photoelectron imaging experiments (see Fig. S1). The detached photoelectrons were accelerated by the stack of electrodes toward a dual microchannel plate (MCP) with an effective diameter of ~40 mm coupled with a phosphor screen (P43) to visualize photoelectron two-dimensional (2D) projections. We used a complementary metal-oxide-semiconductor (CMOS) camera equipped with an event-based vision sensor (SilkyEvCam HD, 1280 × 720 pixels, CenturyArks Co., Ltd.) for recording the photoelectron images, which enables us to perform event counting of bright spots appearing on the phosphor screen. We also recorded an image with a photodetachment laser turned off (i.e., a laser-off image) to obtain a background image of secondary electrons due to the bunched beam of $Ag_{12}W^{-}$ anions itself; some of the anions impinged on the extraction electrodes to generate the secondary electrons, which were detected with the MCP as well. Thus, we subtracted the laser-off image from the laser-on image to obtain a 2D projection of photoelectrons detached from $Ag_{12}W^{-}$ anions, and a total of 55 such images were obtained by varying the laser wavelength. As reported previously[26], we applied 3×3 average filter for the subtracted images to reduce hot spots appearing in the images. The 2D projection thus obtained was analyzed by the polar onion-peeling (POP) method[36] to reconstruct the original 3D distribution of the detached photoelectrons (**Extended Data Fig. 1**), which yielded photoelectron-kinetic-energy (PKE) distributions (i.e., photoelectron spectra, see **Extended Data Fig. 2**) and laboratory-frame photoelectron angular distributions (LF-PADs). The imaging spectrometer was energy-calibrated using the electron affinity of atomic silver (1.303 eV)[37]. The energy resolution was typically 5–6%, e.g., 58 meV for 1 eV photoelectrons.

The amount of $Ag_{12}W^{-}$ ions was also monitored by another CMOS camera (CS165MU1/M, 1440×1080 pixels, Thorlabs) in combination with another set of a MCP and a P43 phosphor screen placed on the exit of the VMI electrodes during photoelectron imaging experiment (see Fig. S1 in the Supplementary Information). The photoelectron yield spectrum was obtained by summing the counts over all pixels in each photoelectron image and normalizing the resulting total by the ion intensity and laser power at each measured wavelength.

Photoelectron imaging and photoelectron yield spectroscopy of $Ag_{12}Mo^{-}$ were also performed using the same procedures as for $Ag_{12}W^{-}$. A total of 59 photoelectron images were recorded at photon energies ranging from 1.23 to 2.89 eV. The bright signal observed at the center of the photoelectron image recorded at 2.89 eV (see **Extended Data Figs. 5 and 6**) may originate from photodetachment of $Ag_{13}^{-}$, because $Ag_{12}Mo^{-}$ and $Ag_{13}^{-}$ could not be fully separated owing to the limited mass resolution. However, as reported previously[27], the photodetachment cross section of $Ag_{13}^{-}$ begins to increase at approximately 2.9 eV. Therefore, the photoelectron yield spectrum of $Ag_{12}Mo^{-}$ shown in **Extended Data Fig. 4A** is free from contributions from $Ag_{13}^{-}$ at photon energies between 1.23 and 2.81 eV.

Photodetachment action spectroscopy

The experimental setup has already been described in our previous reports[38–40]. The overall experimental setup is schematically shown in Fig. S2 in the Supplementary Information. Briefly, a continuous beam of $Ag_{12}W^{-}$ anions was produced by the same procedures as explained above. The $Ag_{12}W^{-}$ anions were delivered into a 40 cm long linear ion trap filled with a buffer helium gas (~$10^{-1}$ Pa), which yielded 0.5–2.0 $\times 10^{8}$ ions stored in the trap. For fine spatial overlap between a laser beam and the ion distribution, the eight pole electrodes were wired to be operated as a quadrupole ion trap, as for the ions to be located near the central axis of the trap[40].

The ions were thermalized for about 500 ms by collisions with the buffer helium gas at room temperature (300 K). The anions in the trap were irradiated with 10 shots of nanosecond laser pulses from an optical parametric oscillator (NT-230-50-SH/SF-DUV, EKSPLA; $\lambda$ = 225–852 nm) operated at a repetition rate of 50 Hz, which was followed by additional 200 ms storage to allow photoexcited $Ag_{12}W^{-*}$ to decay via electron detachment. Residual $Ag_{12}W^{-}$ anions were ejected from the trap to measure their ion currents at a Faraday plate placed near a quadrupole ion deflector (QD3, see Fig. S2) with a picoammeter (Model 6485, Keithley Instruments, Inc.). The number of $Ag_{12}W^{-}$ ions with and without laser irradiation were obtained as $N_{ON}$ and $N_{OFF}$, respectively. The depletion of $Ag_{12}W^{-}$ anions, i.e., $N_{OFF}$–$N_{ON}$, is proportional to a photodetachment cross section, $\sigma_{PD}$, in the linear response regime. Note that dissociation of the photoexcited $Ag_{12}W^{-*}$ can also occur within 200 ms. However, we believe that the depletion signals due to dissociation do not contribute to $N_{OFF}$–$N_{ON}$, as the fragment anions (e.g., $Ag_{11}W^-$) are also detected at the ion detector at QD3 in our setup[27] (see Fig. S2).

To evaluate $\sigma_{PD}$, we first characterized a two-dimensional spatial profile of the ion cloud in the *x*-*y* plane defined in Fig. S2 prior to spectroscopic experiments. The spatial profile was probed in a manner reported in ref[40], where a photodetachment yield of trapped $Ag_{12}W^-$ was measured as a collimated laser beam, with a laser-beam spot size of about 1 mm in diameter, was scanned over the cross section orthogonal to the axis, *z*, of the linear trap (see Fig. S2). The column density of the ions, $n_{ion}(x, y)$, thus obtained showed that the ion cloud typically had a Gaussian-type profile with a full width at half-maximum (FWHM) of about 2.7 mm. On the other hand, the laser beam employed in the spectrum measurements had a nearly flat-top profile with a diameter of about 4 mm. The cross section is then defined as follows:

$$N_{\mathrm{OFF}} - N_{\mathrm{ON}} = \sigma_{\mathrm{PD}} \iint n_{\mathrm{ion}}(x, y) \cdot n_{\mathrm{photon}}(x, y)\, \mathrm{d}x\mathrm{d}y\,, \quad (1)$$

where $n_{photon}(x, y)$ is the column density of photons of the laser beam. The overlap integral can be analytically evaluated for a Gaussian profile of the ions and the flat-top profile of the laser, which derives the following expression for the photodetachment cross section

$$\sigma_{\mathrm{PD}} = \frac{N_{\mathrm{OFF}} - N_{\mathrm{ON}}}{R_{\mathrm{irrad}} \cdot N_{\mathrm{OFF}}} \cdot \frac{A_{\mathrm{laser}}}{N_{\mathrm{photon}}}\,, \quad (2)$$

where $N_{photon}$ and $A_{laser}$ stand for the total number of incident photons and the spot size of the laser beam, respectively. The factor, $R_{irrad}$, represents the fraction of cluster anions irradiated by the laser beam, which was typically about 0.8. We estimate that the absolute values of $\sigma_{PD}$ thus obtained suffer from systematic errors of 30–50%, mainly due to the uncertainty in the overlap between the ions and the laser beam.

Computational details are provided in the Supplementary Information.

## References

1. Cooper, J. W. Photoionization from Outer Atomic Subshells. A Model Study. *Phys. Rev.* **128**, 681–693 (1962).
2. Hudson, R. D. & Carter, V. L. Atomic Absorption Cross Sections of Lithium and Sodium Between 600 and 1000 Å. *J. Opt. Soc. Am.* **57**, 651–654 (1967).
3. Samson, J. A. R. & Stolte, W. C. Precision measurements of the total photoionization cross-sections of He, Ne, Ar, Kr, and Xe. *J. Electron Spectrosc. Relat. Phenom.* **123**, 265–276 (2002).
4. Wahlström, C.-G. *et al.* High-order harmonic generation in rare gases with an intense short-pulse laser. *Phys. Rev. A* **48**, 4709–4720 (1993).
5. Itatani, J. *et al.* Tomographic imaging of molecular orbitals. *Nature* **432**, 867–871 (2004).
6. Carlson, T. A. *et al.* Angle-resolved photoelectron spectroscopy of HCl from a photon energy of 16 to 80 eV. *J. Chem. Phys.* **79**, 2157–2162 (1983).
7. Park, H. & Zare, R. N. Evidence for a Cooper minimum in the photoionization dynamics of the NO D $^2\Sigma^+$ state. *Chem. Phys. Lett.* **225**, 327–334 (1994).
8. Molodtsov, S. L. *et al.* Cooper Minima in the Photoemission Spectra of Solids. *Phys. Rev. Lett.* **85**, 4184–4187 (2000).
9. Hudson, R. D. & Carter, V. L. Absorption of Light by Potassium Vapor between 2856 and 1150 Å. *Phys. Rev.* **139**, A1426–A1428 (1965).
10. Kaiser, H. J., Heinicke, E., Rackwitz, R. & Feldmann, D. Photodetachment measurements of alkali negative ions. *Z. Phys.* **270**, 259–265 (1974).
11. Sandner, W., Gallagher, T. F., Safinya, K. A. & Gounand, F. Photoionization of potassium in the vicinity of the minimum in the cross section. *Phys. Rev. A* **23**, 2732–2735 (1981).
12. Schultze, M. *et al.* Delay in Photoemission. *Science* **328**, 1658–1662 (2010).
13. Alexandridi, C. *et al.* Attosecond photoionization dynamics in the vicinity of the Cooper minima in argon. *Phys. Rev. Res.* **3**, L012012 (2021).
14. Pradhan, G. B. *et al.* Cooper minima: a window on nondipole photoionization at low energy. *J. Phys. B: At. Mol. Opt. Phys.* **44**, 201001 (2011).
15. Ilchen, M. *et al.* Symmetry breakdown of electron emission in extreme ultraviolet photoionization of argon. *Nat Commun* **9**, 4659 (2018).
16. Liang, J. *et al.* Attosecond-resolved non-dipole photoionization dynamics. *Nat. Photon.* **18**, 311–317 (2024).

17. Knight, W. D. *et al.* Electronic Shell Structure and Abundances of Sodium Clusters. *Phys. Rev. Lett.* **52**, 2141–2143 (1984).

18. Mackay, A. L. A dense non-crystallographic packing of equal spheres. *Acta Cryst.* **15**, 916–918 (1962).

19. Pyykkö, P. & Runeberg, N. Icosahedral $WAu_{12}$: A Predicted Closed-Shell Species, Stabilized by Aurophilic Attraction and Relativity and in Accord with the 18-Electron Rule. *Angew. Chem. Int. Ed.* **41**, 2174–2176 (2002).

20. Li, X., Kiran, B., Li, J., Zhai, H.-J. & Wang, L.-S. Experimental Observation and Confirmation of Icosahedral $W@Au_{12}$ and $Mo@Au_{12}$ Molecules. *Angew. Chem. Int. Ed.* **41**, 4786–4789 (2002).

21. Suzuki, Y. *et al.* Photoelectron Imaging Signature for Selective Formation of Icosahedral Anionic Silver Cages Encapsulating Group 5 Elements: $M@Ag_{12}^{-}$ (M = V, Nb, and Ta). *J. Phys. Chem. Lett.* **15**, 4327–4332 (2024).

22. Zhang, M. *et al.* Probing the magnetic and structural properties of the 3d, 4d, 5d impurities encapsulated in an icosahedral $Ag_{12}$ cage. *Phys. B: Condens. Matter* **405**, 642–648 (2010).

23. Tian, Z., Zhang, T., Song, W., Rong, Z. & Song, C. Structures and properties of transition metal-doped silver clusters $M@Ag_{12}$ (M = 3d-5d). *J. Nanoparticle Res.* **27**, 134 (2025).

24. Horio, T. *et al.* Photoelectron imaging of size-selected metal cluster anions in a quasi-continuous mode. *Rev. Sci. Instrum.* **93**, 083302 (2022).

25. Horio, T., Nishizato, T., Suzuki, Y., Matsumoto, K. & Terasaki, A. Anion photoelectron velocity-map imaging using a tunable laser at a 100 kHz repetition rate. *J. Chem. Phys.* **162**, 026101 (2025).

26. Suzuki, Y. *et al.* Photoelectron imaging spectroscopy of $Ag_3^{-}$ in the $S_0$ and $S_1$ states. *J. Chem. Phys.* **162**, 184304 (2025).

27. Kawamura, S. *et al.* Photodestruction Action Spectroscopy of Silver Cluster Anions, $Ag_N^{-}$ ($N$ = 3–19), with a Linear Ion Trap: Observation of Bound Excited States above the Photodetachment Threshold. *J. Phys. Chem. A* **127**, 6063–6070 (2023).

28. Gozem, S. *et al.* Photoelectron Wave Function in Photoionization: Plane Wave or Coulomb Wave? *J. Phys. Chem. Lett.* **6**, 4532–4540 (2015).

29. Gozem, S. & Krylov, A. I. The *ezSpectra* suite: An easy-to-use toolkit for spectroscopy modeling. *WIREs Comput. Mol. Sci.* **12**, e1546 (2022).

30. Bartels, C. *et al.* Probing the Angular Momentum Character of the Valence Orbitals of Free Sodium Nanoclusters. *Science* **323**, 1323–1327 (2009).

31. Moseler, M. *et al.* Thermal effects in the photoelectron spectra of $Na_N^-$ clusters ($N$ = 4–19). *Phys. Rev. B* **68**, 165413 (2003).

32. Aguado, A. & Kostko, O. First-principles determination of the structure of $Na_N$ and $Na_N^-$ clusters with up to 80 atoms. *J. Chem. Phys.* **134**, 164304 (2011).

33. Wigner, E. P. On the Behavior of Cross Sections Near Thresholds. *Phys. Rev.* **73**, 1002–1009 (1948).

34. Sarugaku, S., Arakawa, M., Kawano, T. & Terasaki, A. Electronic and Geometric Effects on Chemical Reactivity of 3d-Transition-Metal-Doped Silver Cluster Cations toward Oxygen Molecules. *J. Phys. Chem. C* **123**, 25890–25897 (2019).

35. Horke, D. A., Roberts, G. M., Lecointre, J. & Verlet, J. R. R. Velocity-map imaging at low extraction fields. *Rev. Sci. Instrum.* **83**, 063101 (2012).

36. Roberts, G. M., Nixon, J. L., Lecointre, J., Wrede, E. & Verlet, J. R. R. Toward real-time charged-particle image reconstruction using polar onion-peeling. *Rev. Sci. Instrum.* **80**, 053104 (2009).

37. Hotop, H., Bennett, R. A. & Lineberger, W. C. Electron affinities of Cu and Ag. *J. Chem. Phys.* **58**, 2373–2378 (1973).

38. Terasaki, A., Majima, T. & Kondow, T. Photon-trap spectroscopy of mass-selected ions in an ion trap: Optical absorption and magneto-optical effects. *J. Chem. Phys.* **127**, 231101 (2007).

39. Terasaki, A., Majima, T., Kasai, C. & Kondow, T. Photon-trap spectroscopy of size-selected free cluster ions: "direct" measurement of optical absorption of $Ag_9^+$. *Eur. Phys. J. D* **52**, 43–46 (2009).

40. Majima, T. *et al.* Spatial distribution of ions in a linear octopole radio-frequency ion trap in the space-charge limit. *Phys. Rev. A* **85**, 053414 (2012).

## Acknowledgements

The authors wish to acknowledge Y. Ushiki, Y. Aoki, Y. Kusaba, M. Shimizu, and Y. Toyama for their experimental assistance. The computational work was carried out by the computer facilities at the Research Institute for Information Technology, Kyushu University and the Research Center for Computational Science, Okazaki, Japan (Project: 24-IMS-C215, 26-IMS-C130). Generative AI was used to assist with the translation of portions of the manuscript from Japanese into English. The AI-generated text was subsequently reviewed and edited by the authors. The authors also thank FORTE Science Communications (https://www.forte-science.co.jp/) for English language editing. The authors take full responsibility for the content of the manuscript.

**Author information**

Authors and Affiliations

**Department of Chemistry, Faculty of Science, Kyushu University; 744 Motooka, Nishi-ku, Fukuoka 819-0395, Japan**

Y. Suzuki, M. Tanaka, A. Terasaki, T. Horio

**School of Medical Technology, Health Sciences University of Hokkaido; 1757 Kanazawa, Tobetsucho, Ishikari, Hokkaido 061-0293, Japan**

Y.-I. Suzuki

**Quantum and Spacetime Research Institute (QuaSR), Kyushu University, 744 Motooka, Nishi-Ku, Fukuoka 819-0395, Japan**

A. Terasaki

Contributions

Conceptualization: Y. S. and T. H. Methodology: Y. S., A. T., and T. H. Validation: Y. S. and T. H. Formal analysis: Y. S. Investigation: Y. S. and M. T. Visualization: Y. S. Supervision: Y.-I. S. and A. T. Project administration: T. H. Writing (original draft): Y. S. and T. H. Writing (review

& editing): Y. S., M. T., Y.-I. S., A. T., and T. H. All authors discussed the results and commented on the manuscript.

Corresponding authors

Correspondence to T. Horio and A. Terasaki.

**Ethics declarations**

Competing interest

The authors declare no competing interests.

**Funding**

This work was supported by Grants-in-Aid for Scientific Research (A) (Grant No. JP22H00317), Scientific Research (B) (Grant Nos. JP22H02036, JP23K23304, and JP25K01730), Challenging Research (Exploratory) (Grant No. JP22K19009), Challenging Research (Pioneering) (Grant No. JP23K17349), and JSPS Fellows (Grant No. JP26KJ1781) from the Japan Society for the Promotion of Science (JSPS), the Matsuo Foundation, the Shimadzu Science Foundation, and the Quantum and Spacetime Research Institute, Kyushu University.

**Extended Data**

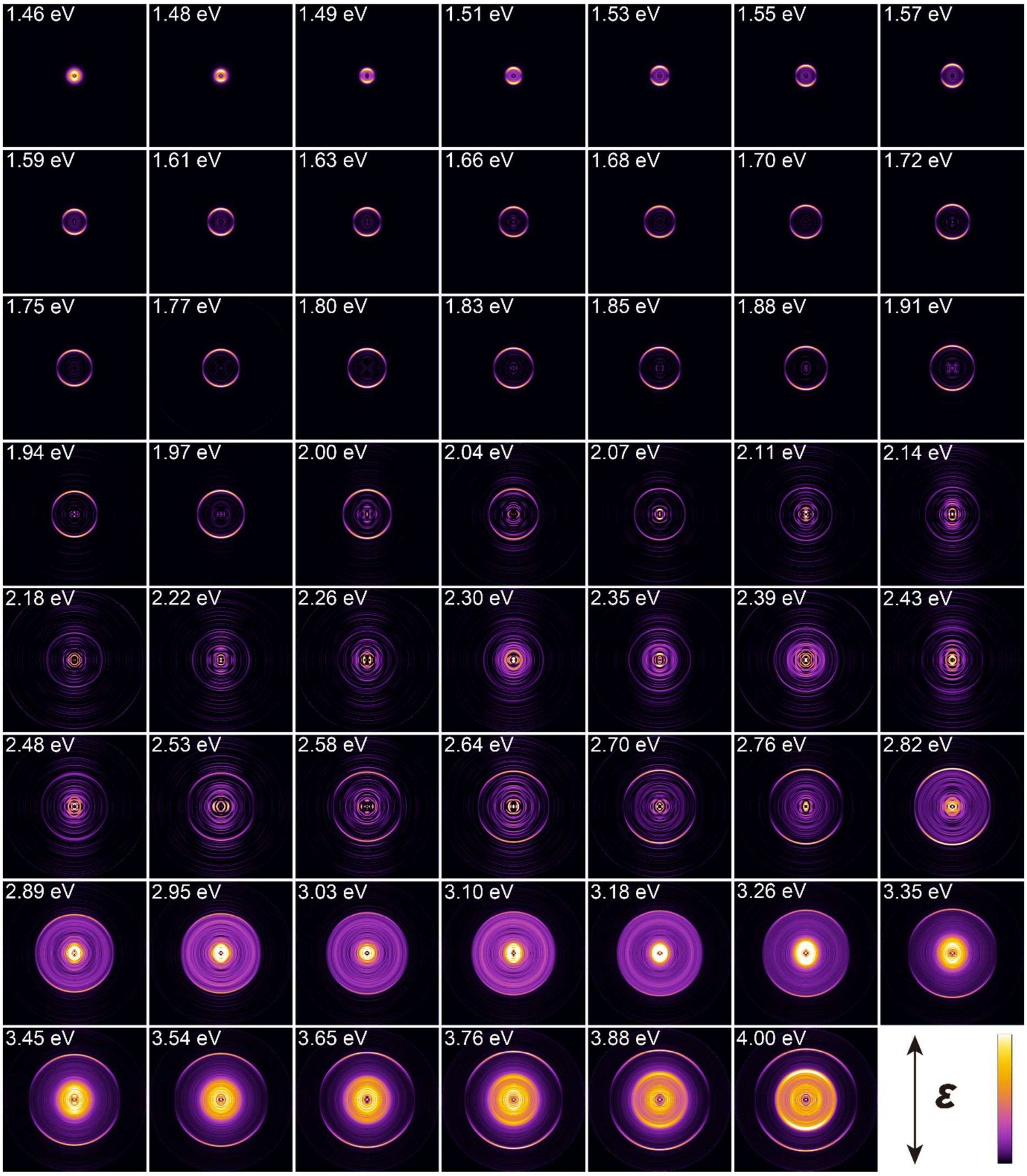

**Fig. 1. Photoelectron slice images of $Ag_{12}W^{-}$.**
55 photoelectron slice images of $Ag_{12}W^{-}$ are recorded at photon energies from 1.46 to 4.00 eV. The polarization vector, *ε*, of the photodetachment laser is indicated by a double arrow (lower right corner). The intensity scale of each slice image was adjusted independently to enhance the visibility of the photoelectron signals. Thus, the intensities should not be quantitatively compared among the images; they do not represent the relative photodetachment cross sections.

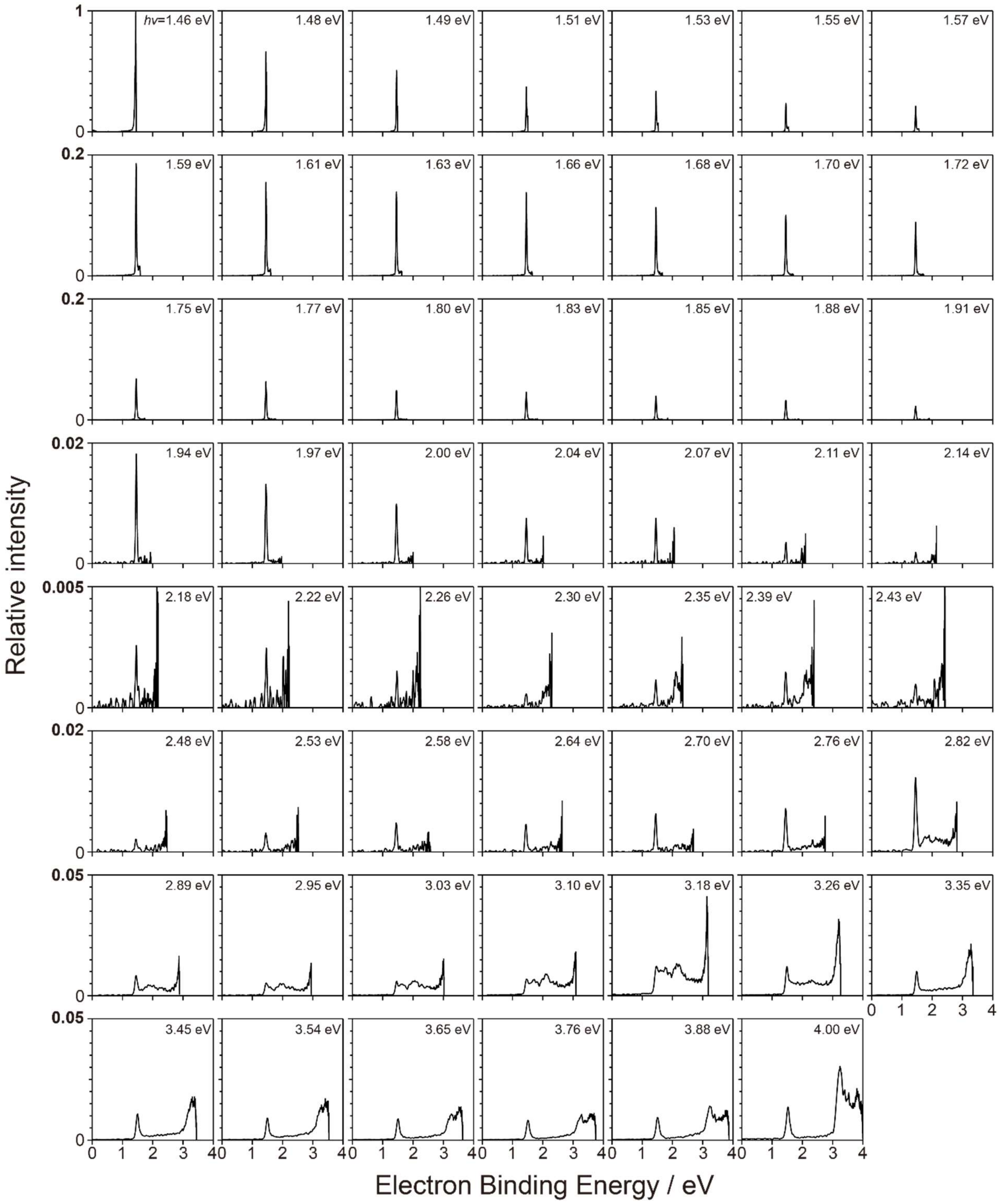

**Fig. 2. Photoelectron spectra of $Ag_{12}W^{-}$.**
Photoelectron spectra of $Ag_{12}W^{-}$ are shown for photon energies from 1.46 to 4.00 eV. The integrated intensity of each spectrum is proportional to the photodetachment cross section (shown in Fig. 4A). Within each row, all spectra are displayed on the same intensity scale as the leftmost spectrum.

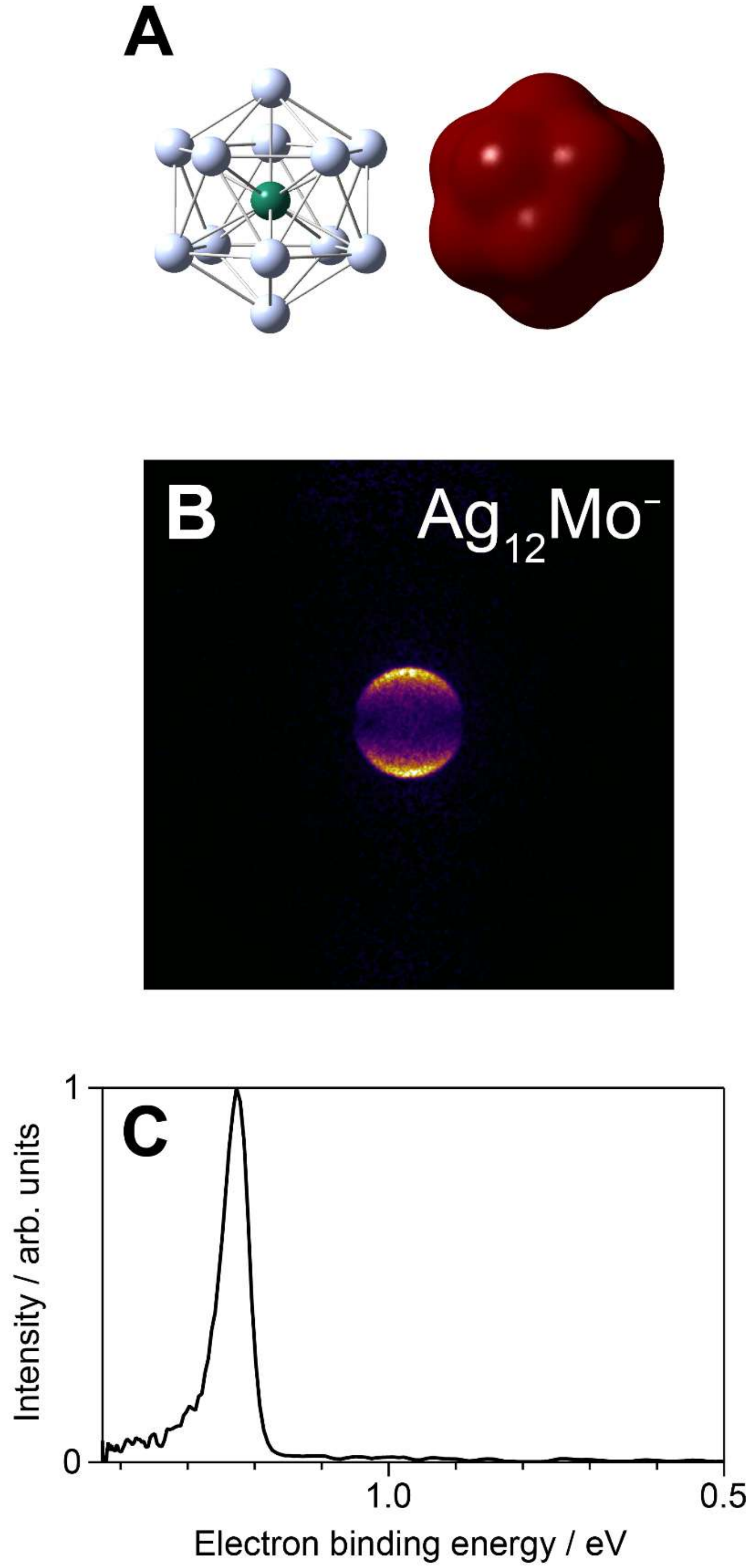

**Fig. 3. Laboratory-frame photoelectron angular distribution and photoelectron spectrum reflect the atom-like character of $Ag_{12}Mo^-$ upon photodetachment from SOMO.**
(A) Optimized geometry of $Ag_{12}Mo^-$ and the isosurface plot of its SOMO (S-like orbital) obtained at PBEPBE/LANL2DZ level of theory (isovalue = 0.010 a.u.). Ag and Mo atoms are displayed as white and green spheres, respectively. (B) Photoelectron image of $Ag_{12}Mo^-$ obtained at a photon energy of 1.43 eV, and (C) the corresponding photoelectron spectrum.

.

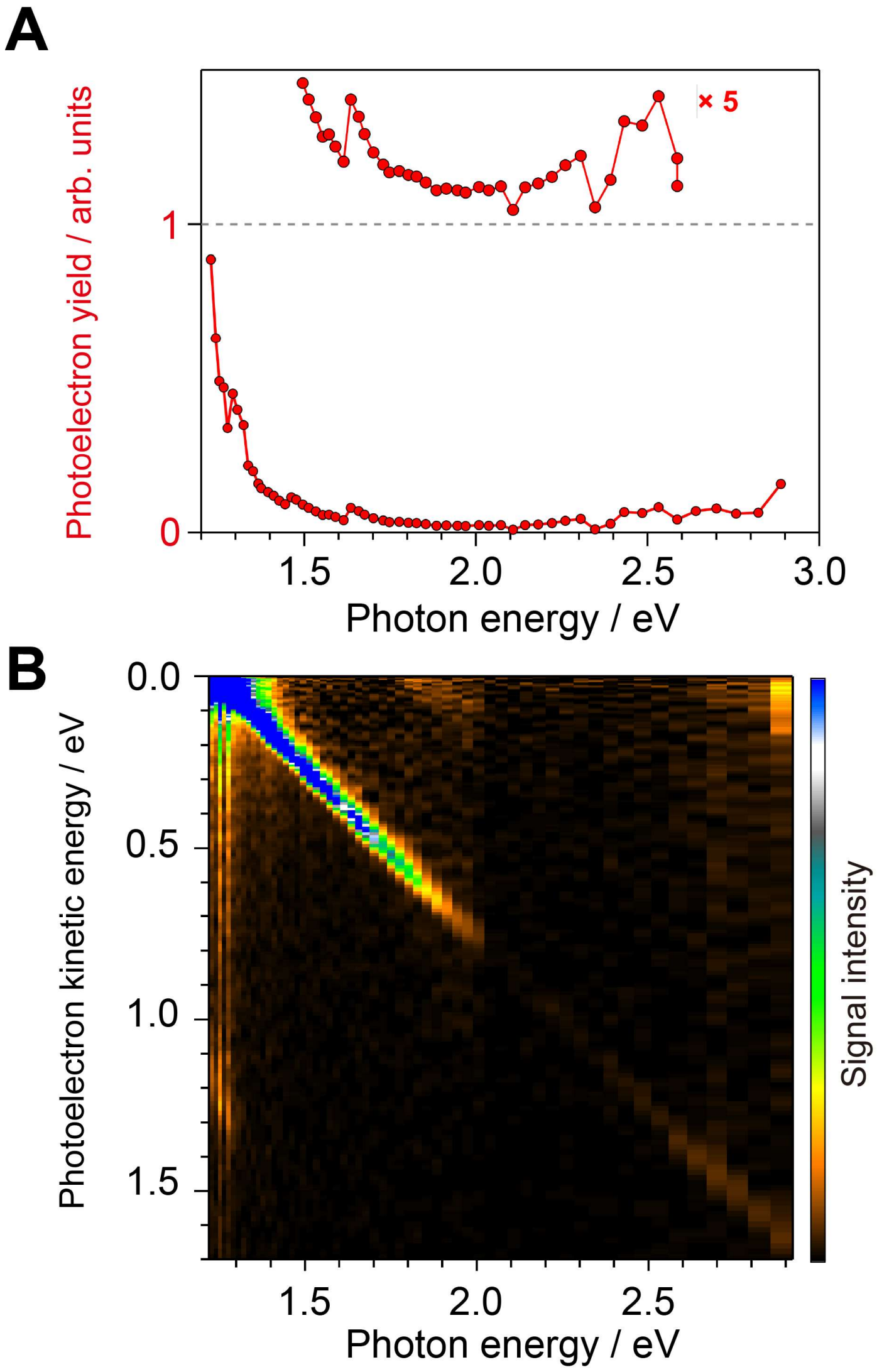

**Fig. 4. Characterization of the zero photoemission from $Ag_{12}Mo^{-}$ by photoelectron yield spectroscopy and photoelectron imaging spectroscopy.**
(A) Experimental photoelectron yield spectrum as a function of photon energy. The inset shows a 5 × enlarged view of the experimental photoelectron yield. (B) Two-dimensional map of photoelectron spectra versus photon energy; the integrated photoelectron spectrum at each photon energy is proportional to the red circles shown in (A).

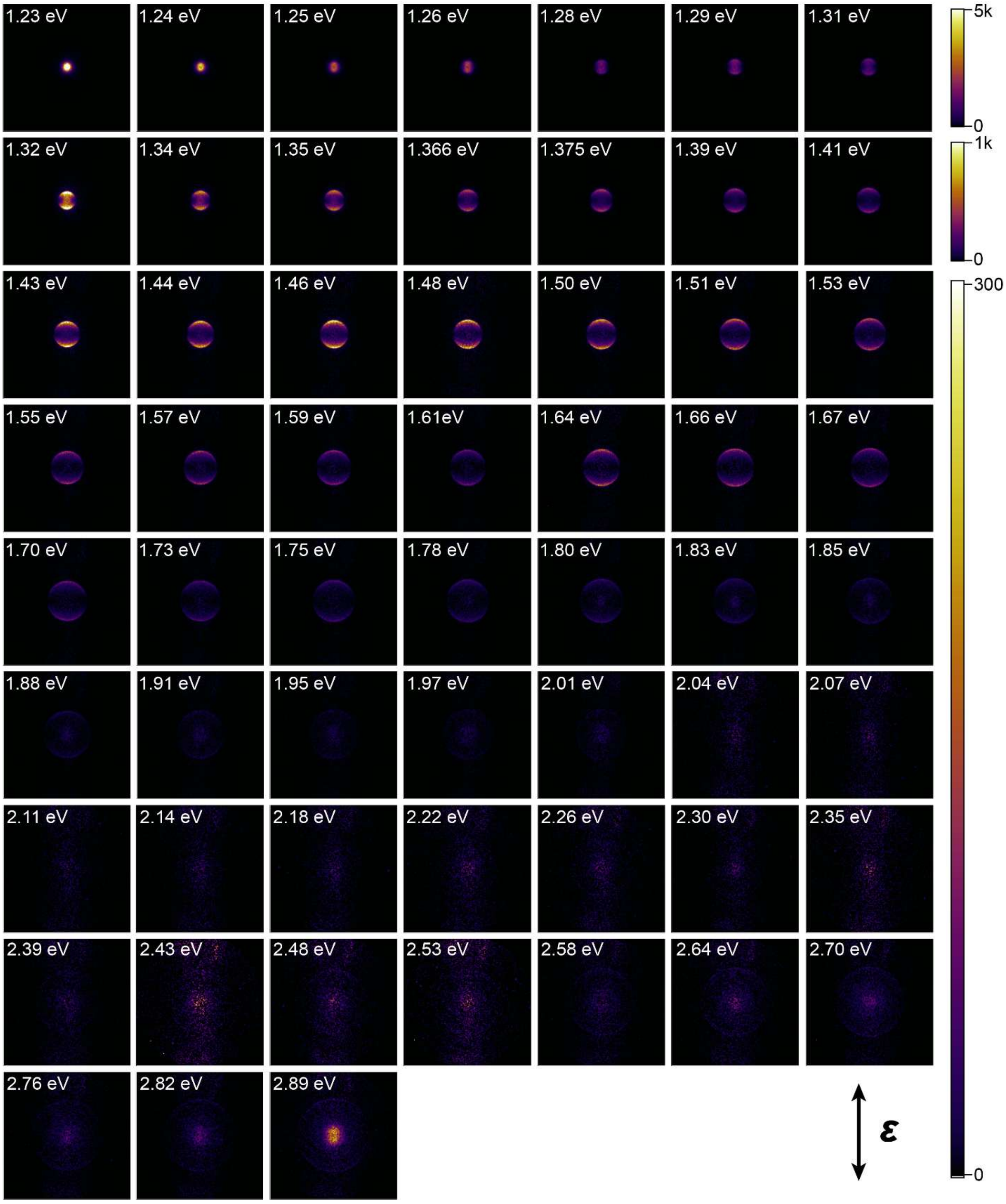

**Fig. 5. Photoelectron images (2D projections) of $Ag_{12}Mo^-$.**
Photoelectron images (2D projections) of $Ag_{12}Mo^-$ are obtained at photon energies from 1.23 to 2.89 eV. The polarization vector, *ε*, of the photodetachment laser is indicated by a double arrow in the bottom-right corner. The total photoelectron counts for each image were normalized by the total number of photons in the incident laser beam and the anion beam intensity; they therefore reflect the relative photodetachment yield (Extended Data Fig. 4A). The color scales of images in the first, second, third, and subsequent rows are shown on the right. The corresponding slice images are provided in Extended Data Fig. 6.

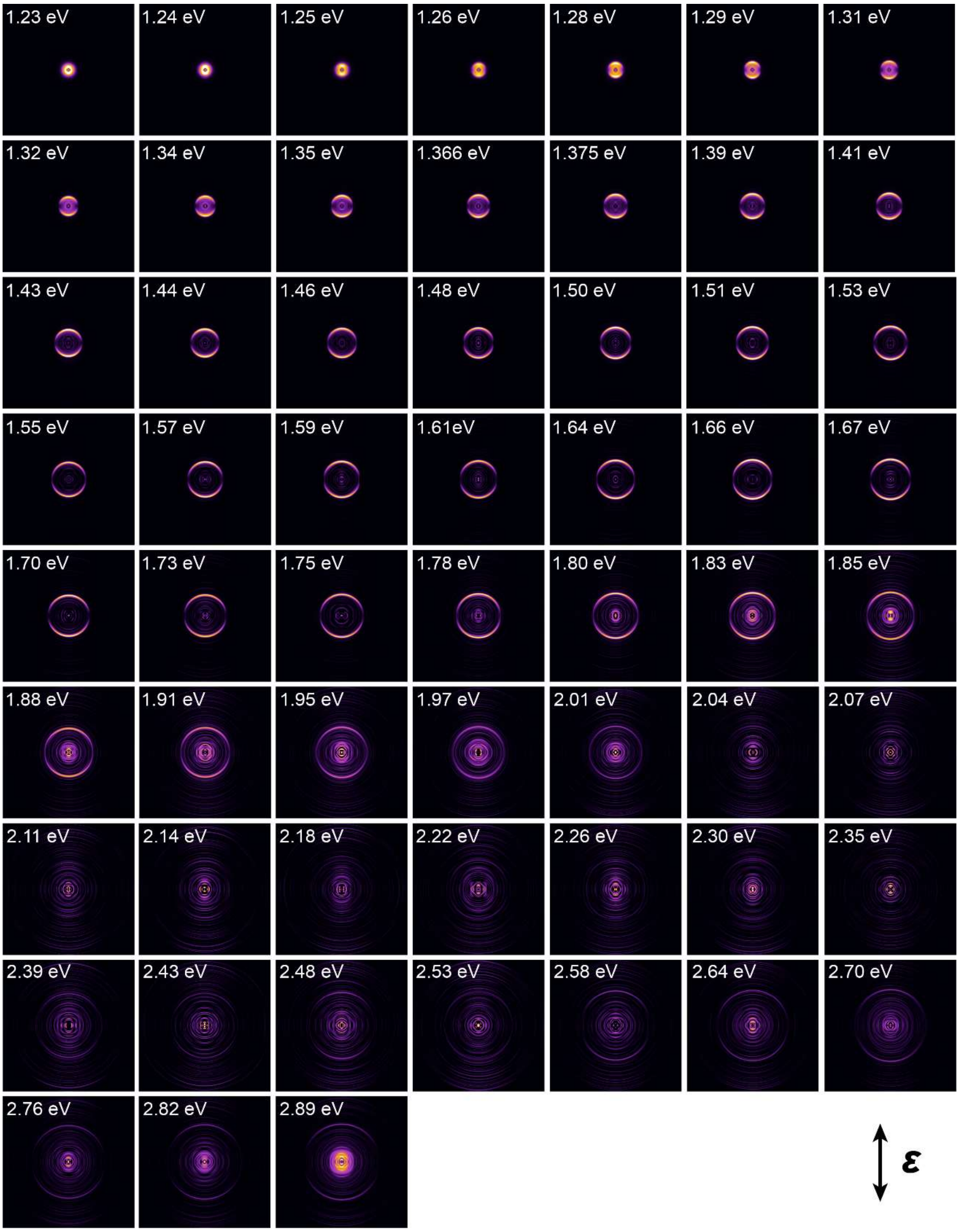

**Fig. 6. Photoelectron slice images of $Ag_{12}Mo^{-}$.**
59 photoelectron slice images of $Ag_{12}Mo^{-}$ are recorded at photon energies from 1.23 to 2.89 eV. The polarization vector, $\boldsymbol{\varepsilon}$, of the photodetachment laser is indicated by a double arrow (bottom-right corner). The intensity scale of each slice image was adjusted independently to enhance the visibility of the photoelectron signals. Thus, the intensities should not be quantitatively compared among the images; they do not represent the relative photodetachment cross sections.

# Supplementary Information for
# "Zero photoemission from icosahedral superatoms"

Yuta Suzuki, Mamoru Tanaka, Yoshi-ichi Suzuki, Akira Terasaki, and Takuya Horio

Correspondence to: terasaki@chem.kyushu-univ.jp, horio@chem.kyushu-univ.jp

**This PDF file includes:**

Sections S1 to S3

S1. Density functional theory (DFT) and time-dependent DFT (TD-DFT) calculations

S2. Simulating photoelectron anisotropy parameter and photodetachment cross sections

S3. Calculation of the radial part of the transition dipole moments as a function of the charge for a Coulomb wave and photoelectron kinetic energy

Figures S1 to S10

Tables S1 to S2

**S1. Density functional theory (DFT) and time-dependent DFT (TD-DFT) calculations**

We performed DFT calculations to explore the most stable geometries of $Ag_{12}W^{-}$, starting from 210 initial geometries of endohedrally or exohedrally doped structures. The spin multiplicity was restricted to doublet and quartet. The geometry optimizations were conducted by the Gaussian16 package[1] using possible combinations of BP86[2,3], PBEPBE[4], PBE1PBE[5], and HSEH1PBE[6] functionals with LANL2DZ[7] and SDD[8,9] basis sets for both Ag and W atoms with the corresponding effective core potentials (ECPs). Frequency analyses were also performed to confirm that the obtained structures are in true local minima. Across all combinations of functionals and basis sets, the $I_h$ form is found to be the most stable structure (Fig. S3). Then, we refined the geometry optimization of the $I_h$ structure with all possible combinations of BP86, PBEPBE, PBE1PBE, HSEH1PBE, B3LYP[10–12] functionals with LANL2DZ, SDD, def2-SVPD[8,13,14], def2-TZVPP[8,13], def2-TZVPD[8,13,14], cc-pVDZ-PP[15–17], and aug-cc-pVDZ-PP[15–17].

In addition, to assign the photoelectron spectrum, VDE of $Ag_{12}W^{-}$ was calculated from the total energy difference between optimized $Ag_{12}W^{-}$ and the corresponding neutral $Ag_{12}W$ having exactly the same geometry as optimized for $Ag_{12}W^{-}$. (Table S2)

TD-DFT calculations were also carried out to assign the photodetachment action spectrum and photoelectron yield spectrum. Vertical excitation energies (VEEs) are calculated by the combination of HSEH1PBE functional and LANL2DZ, SDD, def2-SVP[8,13], def2-SVPD, def2-TZVP[8,13], def2-TZVPD, cc-pVDZ-PP, cc-pVTZ-PP[15–17]. The spectra are simulated by convolution of the calculated stick spectra (oscillator strength) with a Gaussian functions with FWHM of 0.20 eV (Fig. S6).

Furthermore, we also performed geometry optimization of neutral $Ag_{12}W$ for the calculation of adiabatic detachment energy (ADE). For geometry optimization, BP86, PBEPBE, PBE1PBE, and HSEH1PBE functionals with LANL2DZ were examined. The spin multiplicity was restricted to singlet and triplet (The result is shown in the Fig. S4). ADEs were calculated by the total energy difference between optimized $Ag_{12}W^{-}$ and optimized $Ag_{12}W$.

We also optimized the geometries of $Ag_{12}Mo^{-}$ and $Ag_{12}Mo$ at the PBEPBE/LANL2DZ level of theory, which provided the best agreement between the experimental and calculated VDEs for $Ag_{12}W^{-}$. Both doublet and quartet spin states were considered for $Ag_{12}Mo^{-}$, whereas singlet and triplet spin states for neutral $Ag_{12}Mo$. The optimized geometries are summarized in Fig. S10. The calculated VDE and ADE are both 1.18 eV, in good agreement with the experimental VDE of 1.23 eV. In addition, the SOMO of $Ag_{12}Mo^{-}$ was found to be the 2S-like orbital, as shown in **Extended Data Fig. 3A** of the main text. This similarity supports the observed strongly positive photoelectron angular anisotropy, $\beta \sim 2$ of $Ag_{12}Mo^{-}$ as displayed in Fig. S9.

**S2. Simulating photoelectron anisotropy parameter and photodetachment cross sections**

Although Dyson orbitals are required for rigorously evaluating matrix elements of photodetachment processes, their computations are not feasible for the present system. Instead, we substituted Kohn−Sham orbitals for Dyson orbitals. We used Q-Chem 5.4 package[18] to perform DFT calculations with PBE functional and LANL2DZ basis set with the corresponding ECP as used for the geometry optimization with Gaussian16 to obtain Kohn-Sham orbitals of $Ag_{12}W^{-}$. These orbitals were used to calculate the photoelectron angular anisotropy parameter, $\beta$, and photodetachment cross section by ezDyson 5.0[19,20]. In this calculation, we assumed a coulomb wave with +0.37 *e* charge for photoelectron wavefunction to account for the Cooper minimum observed around 2.3 eV in photon energy (see the main text); we truncated $l_{max}$, the maximum number of the orbital angular momentum, at 7 (i.e., j-wave) in the expansion of spherical harmonics.

Figs. S5A–D show the calculated results for the $S_0(^1A_g) + e^- \leftarrow D_0(^2A_g)$ transition. Figs. S5A and S5B show the photodetachment cross section and the absolute squared values of the coefficients for the partial wave expansion, $|C_{kl}|^2$, respectively; the latters represent the contribution of each outgoing partial wave with orbital angular momentum quantum number $l$. In the ezDyson program, the $m$-resolved quantities $|C_{klm}|^2$ are calculated for each magnetic quantum number $m$. We therefore summed $|C_{klm}|^2$ over all allowed $m$ values for each $l$ to obtain $|C_{kl}|^2$. The calculated cross section is determined by the sum of these contributions over $l$, together with the $k$- and photon-energy-dependent prefactors included in the ezDyson program, where photoelectron kinetic energy $E_k = k^2 / 2$ in atomic units. Fig. S5C shows the relative contribution of each partial wave, obtained by normalizing each $|C_{kl}|^2$ by the sum over all $l$ values. Fig. S5D shows the calculated anisotropy parameter $\beta$. Figs. S5E–H show the corresponding calculated results for the $S_1(^1H_g) + e^- \leftarrow D_0(^2A_g)$ transition channel, corresponding to Figs. S5A–D, respectively.

**S3. Calculation of the radial part of the transition dipole moments as a function of the charge for a Coulomb wave and photoelectron kinetic energy**

Investigation of the radial part of the transition dipole moment, $m_r(Z_{eff}, E_k)$, as a function of the charge for a Coulomb wave ($Z_{eff}$) and photoelectron kinetic energy ($E_k$), helps us to further discuss the Cooper minimum discovered in the present experiments. We numerically evaluated $m_r(Z_{eff}, E_k)$ for the $S_0(^1A_g) + e^- \leftarrow D_0(^2A_g)$ transition using a home-built Python code. As with the calculations of $\beta(E_k)$ and the photodetachment cross section, $\sigma(E_k)$, using ezDyson, the Kohn-Sham SOMO of $Ag_{12}W^-$ obtained at the PBEPBE/LANL2DZ level of theory was used instead of the Dyson orbital. The orbital values were extracted from a cube (.cub) file generated by Gaussian16. The detached photoelectron was described by a Coulomb wave with $l = 1$ (p-wave). The numerical integration for evaluating $m_r(Z_{eff}, E_k)$ was carried out by varying $Z_{eff}$ from 0 to $+1.00e$ in increments of $0.01e$, and $E_k$ from 0 to 1.54 eV in steps of 0.005 eV. In this manner, a two-dimensional map of $|m_r(Z_{eff}, E_k)|^2$ was obtained, which is shown in Fig. S7. Fig. S8 compares a calculated two-dimensional map of the photodetachment cross-sections, obtained by Eq. (1), with the experimental photodetachment action spectrum measured in the vicinity of the Cooper minimum. In Eq. (1), $\sigma$ and $c$ denote the cross section and the speed of light, respectively.

$$\sigma = \frac{8\pi kE}{c} |m_r(Z_{eff}, E_k)|^2 \tag{1}$$

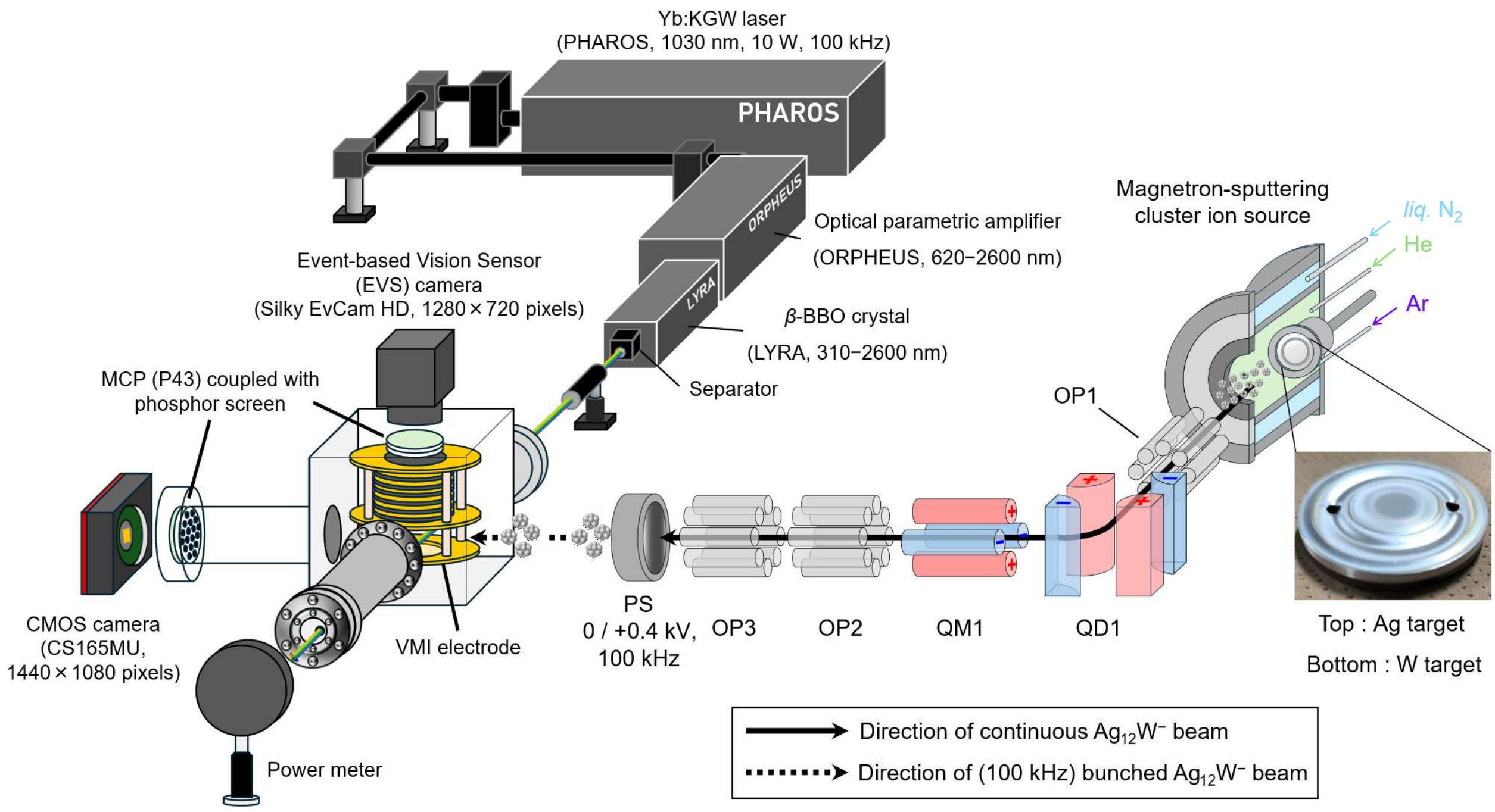

**Fig. S1. Schematic illustration of the apparatus for photoelectron imaging spectroscopy and photoelectron yield spectroscopy.**

Abbreviations: OP, Octupole ion guide; QD, Quadrupole ion deflector; QM, Quadrupole mass filter; PS, Potential switch.

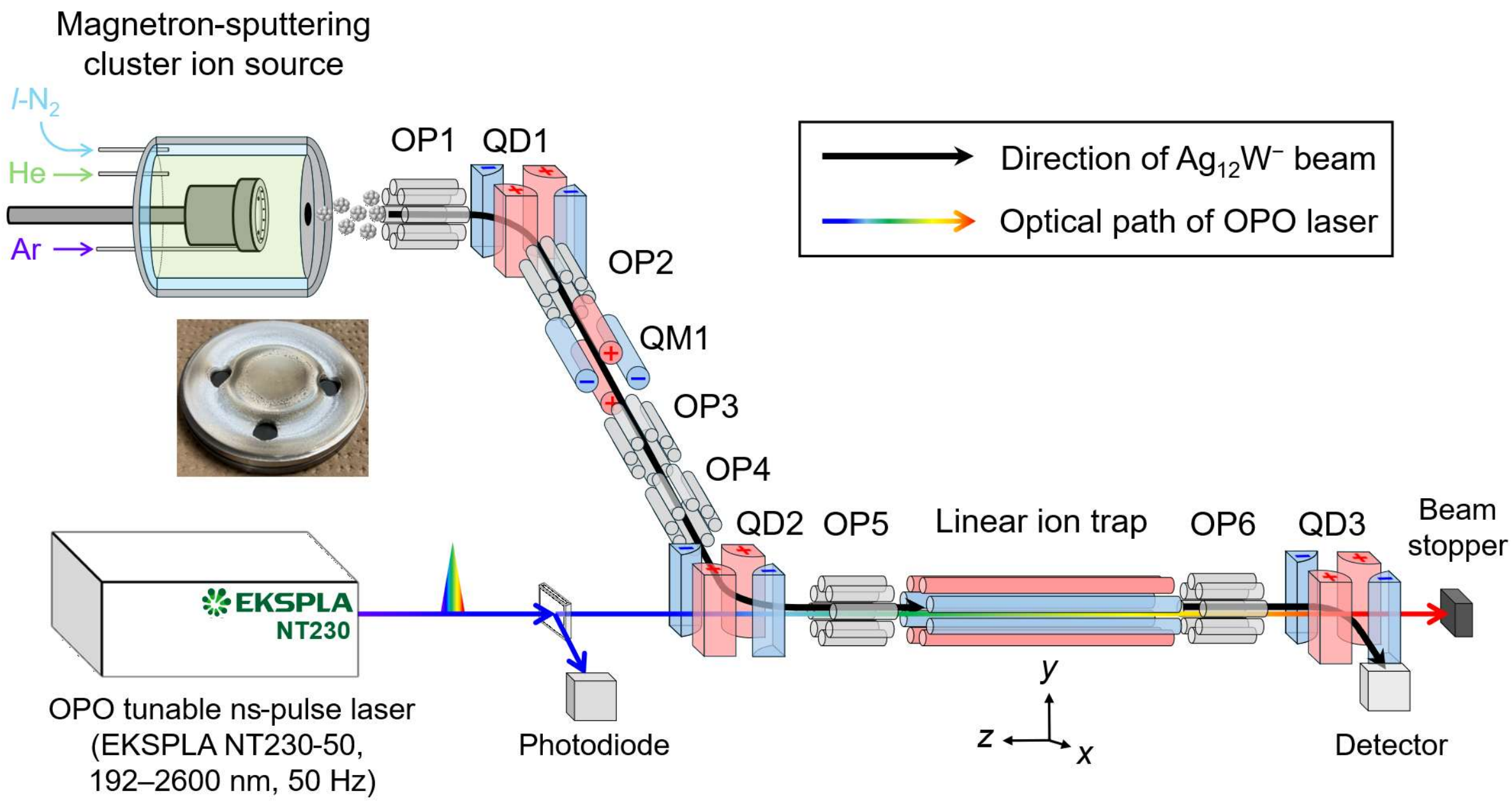

**Fig. S2. Schematic illustration of the apparatus for photodetachment action spectroscopy.** Abbreviations: OP, Octupole ion guide; QD, Quadrupole ion deflector; QM, Quadrupole mass filter; OPO, Optical parametric oscillator.

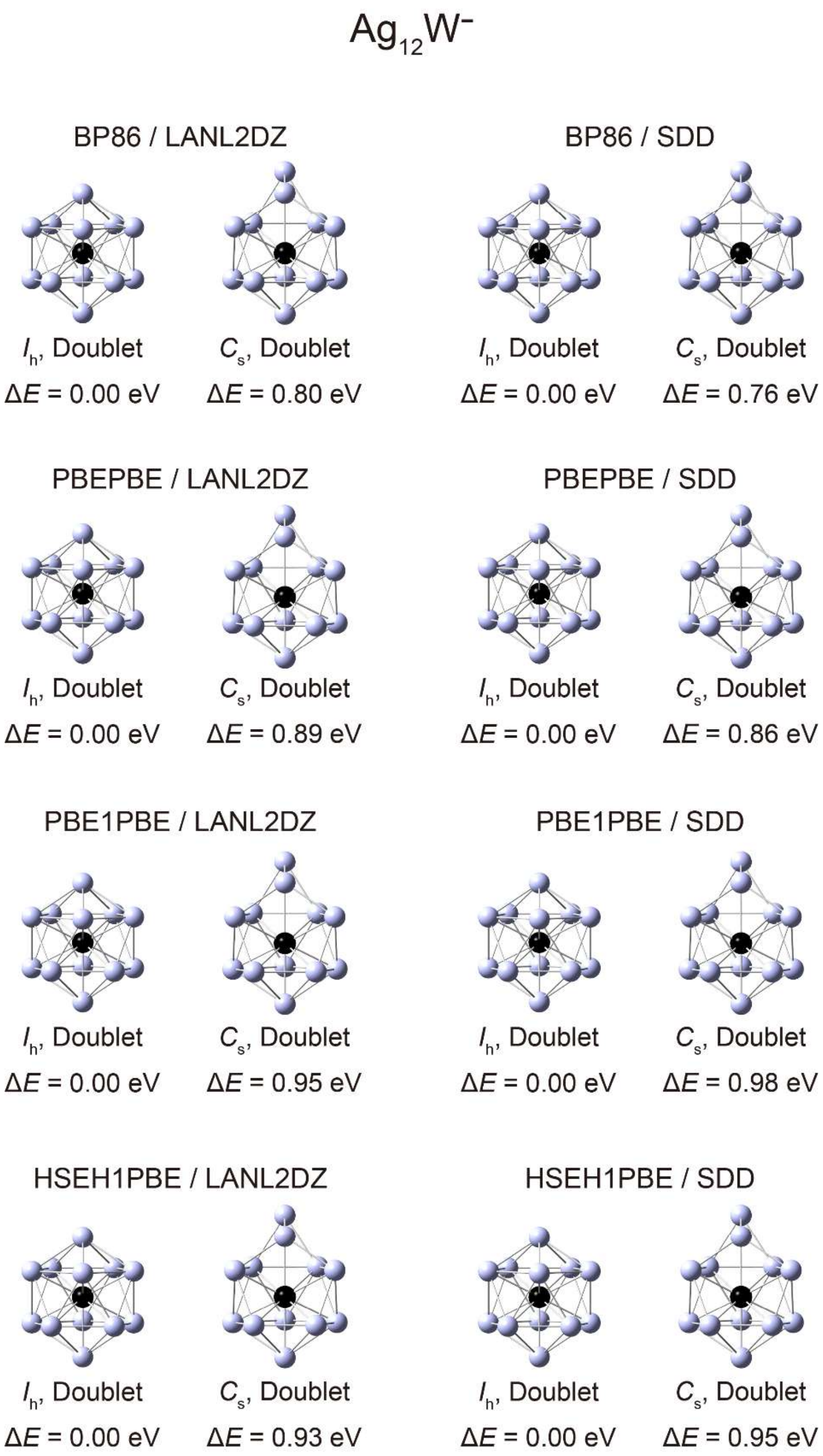

**Fig. S3. Optimized geometric structures of $Ag_{12}W^{-}$.**

The lowest and the second lowest energy isomers of anionic $Ag_{12}W^{-}$ at eight computational levels. Their symmetry, spin state, and relative energy ($\Delta E$) are also presented below each structure.

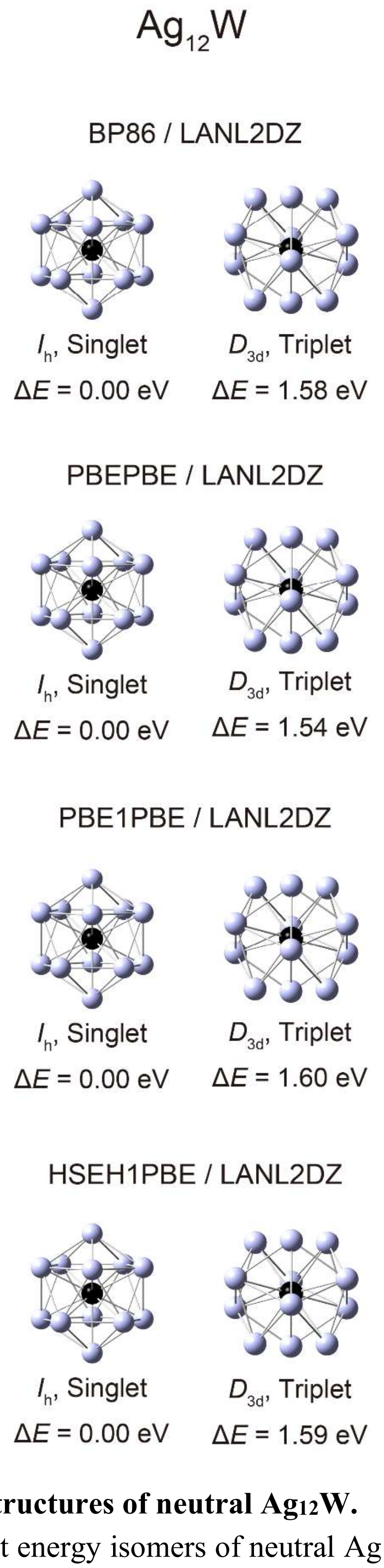

**Fig. S4. Optimized geometric structures of neutral $Ag_{12}W$.**
The lowest and the second lowest energy isomers of neutral $Ag_{12}W$ at four computational levels. Their symmetry, spin state, and relative energy ($\Delta E$) are also presented below each structure.

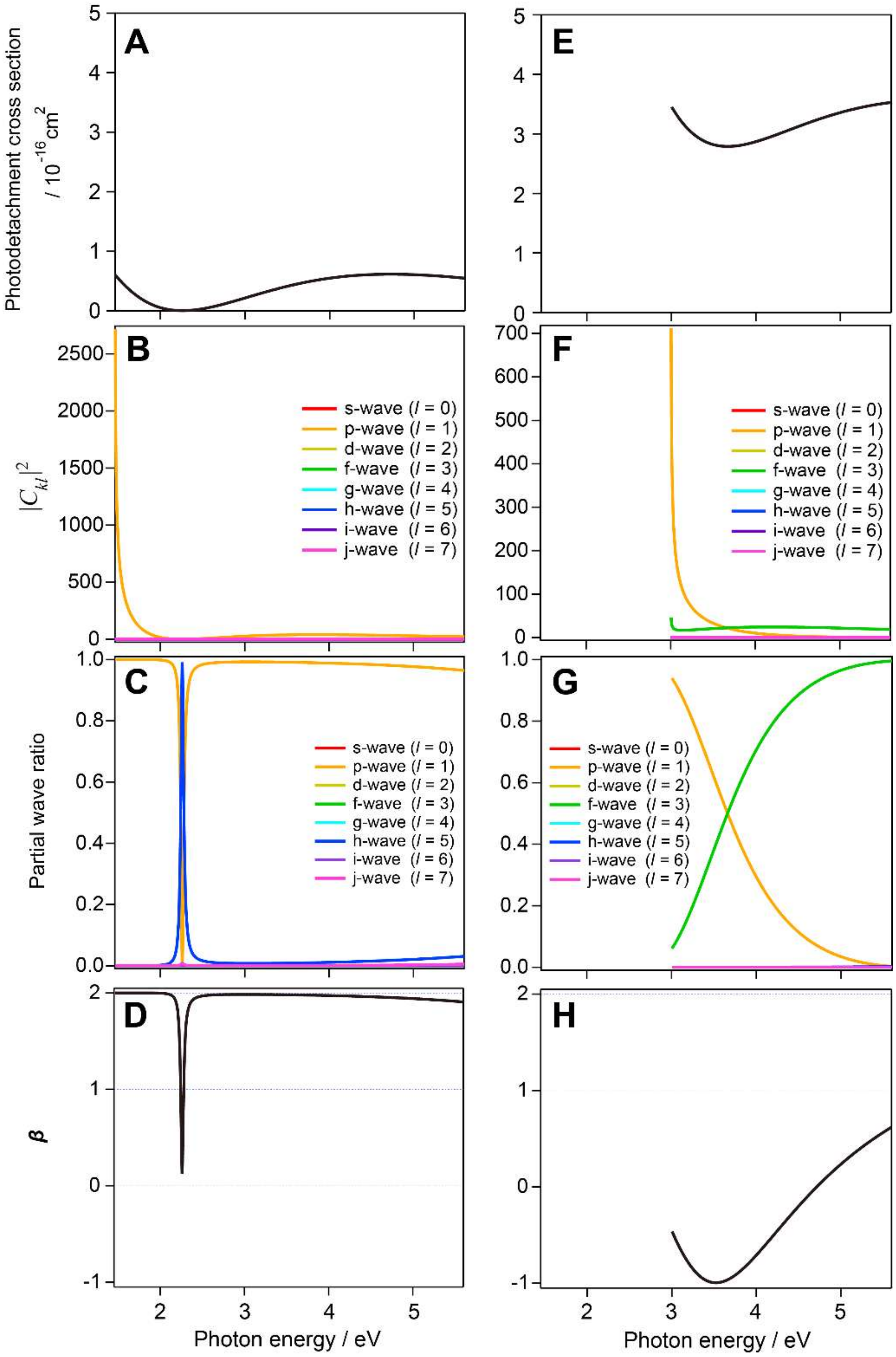

**Fig. S5. Calculated results of ezDyson for $Ag_{12}W^-$.**

(A) Photodetachment cross section as a function of photon energy between 1.46 and 5.6 eV, (B) amplitude of each spherical wave squared (i.e., $|C_{kl}|^2$), (C) partial wave ratio, (D) $\beta$ parameter simulated by ezDyson at PBEPBE/LANL2DZ level of theory for $S_0(^1A_g) + e^- \leftarrow D_0(^2A_g)$ transition (i.e., photodetachment from S-like ($a_g$) orbital). (E), (F), (G), and (H) correspond to (A), (B), (C), and (D), respectively, but for $S_1(^1H_g) + e^- \leftarrow D_0(^2A_g)$ transition (i.e., photodetachment from one of the D-like ($h_g$) orbitals).

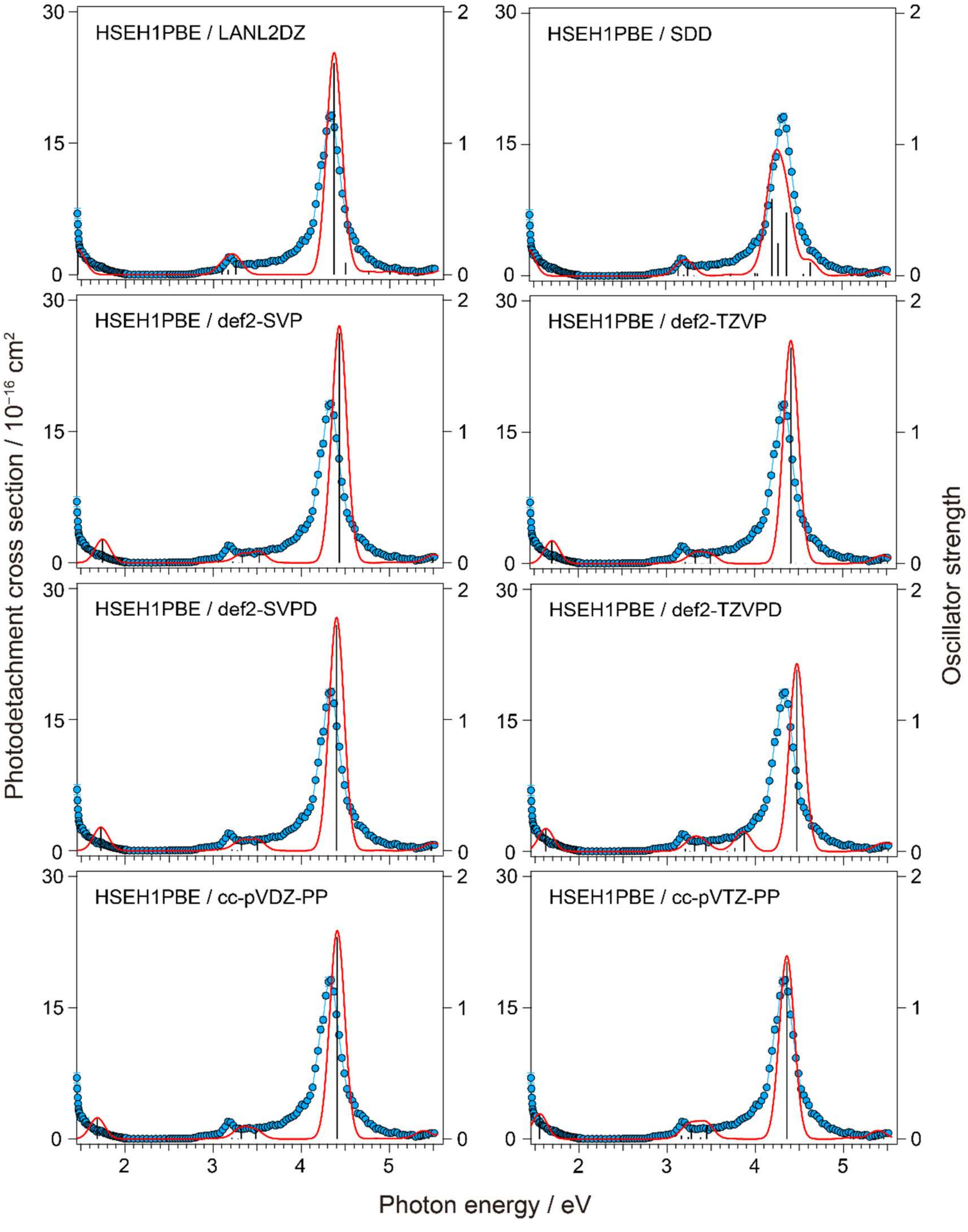

**Fig. S6. TD-DFT results of $Ag_{12}W^-$.**

Comparison of the experimental photodetachment action spectra (skyblue marker) with the simulated spectra by TD-DFT calculations (red solid line), which are obtained by convolution of the calculated stick spectra (oscillator strength) with a Gaussian function with FWHM of 0.20 eV.

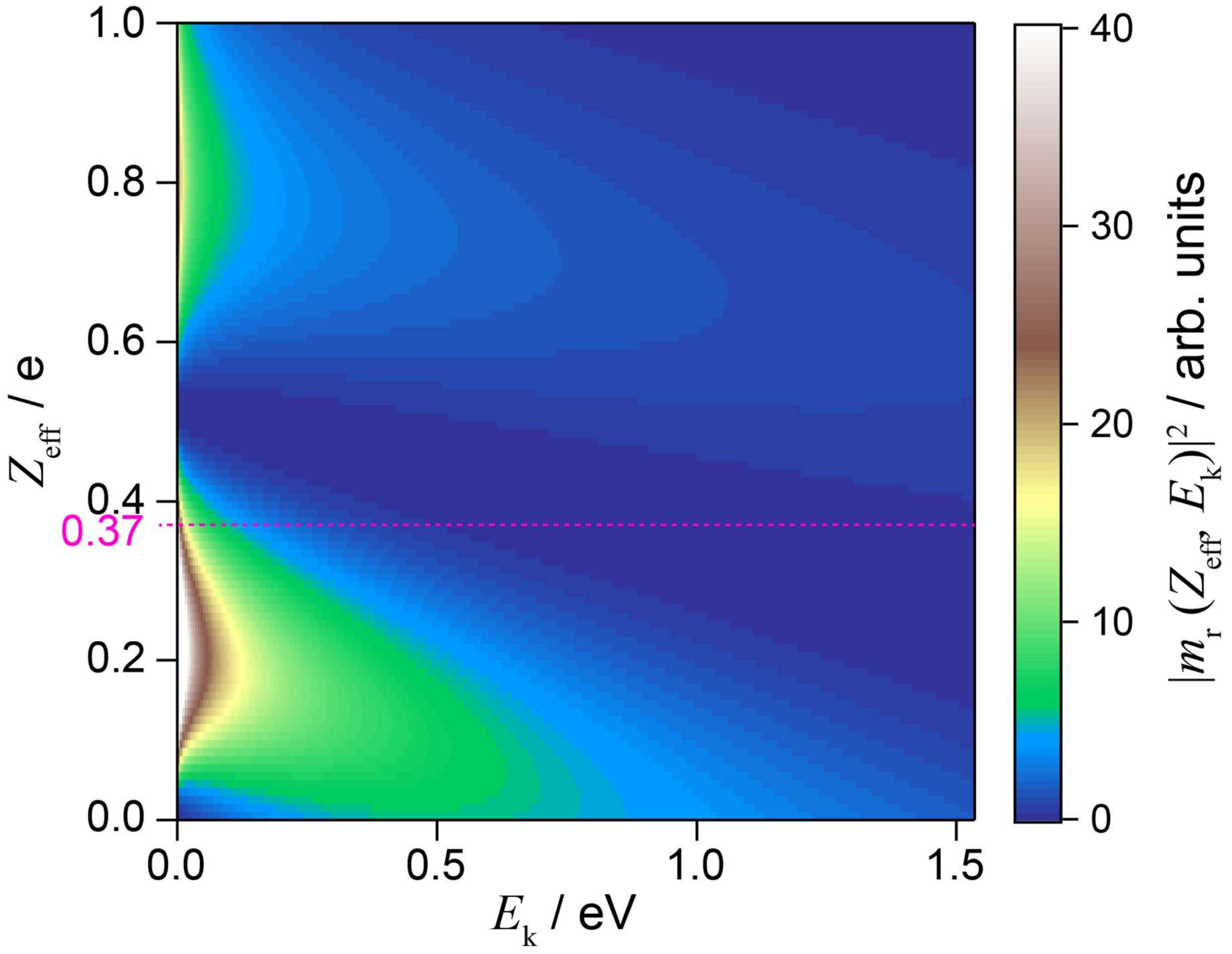

**Fig. S7. Computational results of the squared transition dipole moment of $Ag_{12}W^-$.**
The 2D map of the calculated $|m_r(Z_{eff}, E_k)|^2$ as a function of the charge of the residual core $Z_{eff}$ and the photoelectron kinetic energy $E_k$. The pink horizontal dashed line corresponds to $Z_{eff} = 0.37e$ that best reproduces the experimental photodetachment cross section (see Fig. S10).

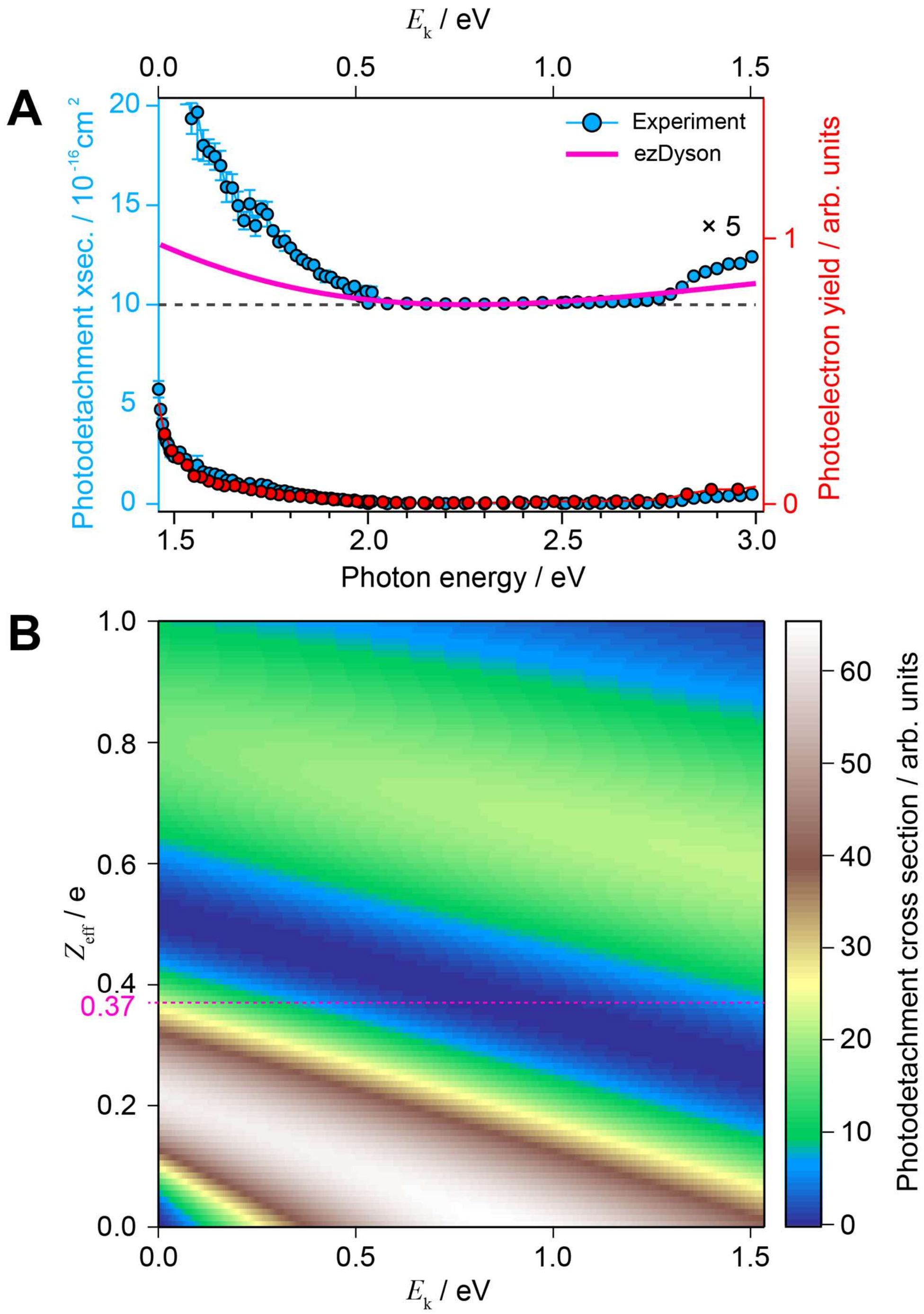

**Fig. S8. Experiment vs. calculated photodetachment cross sections of $Ag_{12}W^{-}$.**

(A) Photodetachment action spectra and photoelectron yield spectra as a function of photon energy from 1.46 to 3.00 eV, which corresponds to the $E_k$ up to 1.54 eV. (B) The 2D map of the photodetachment cross section with respect to the effective charge of the residual core $Z_{eff}$ and the photoelectron kinetic energy $E_k$. The pink horizontal dashed line corresponds to $Z_{eff} = 0.37e$.

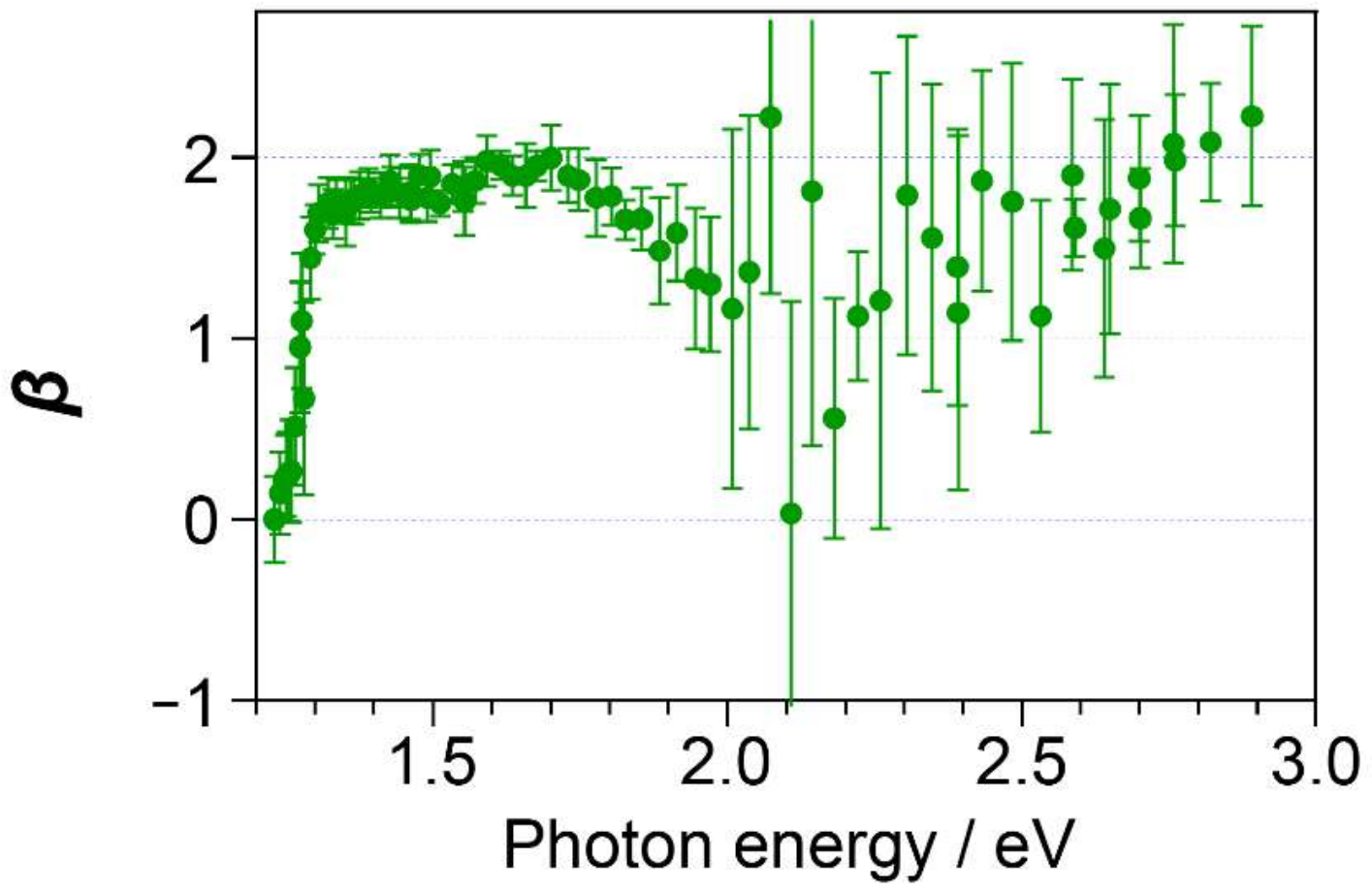

**Fig. S9. *β* plot of $Ag_{12}Mo^-$.**

Photoelectron anisotropy parameters, *β*, for the $S_0(^1A_g) + e^- \leftarrow D_0(^2A_g)$ transition of $Ag_{12}Mo^-$ as a function of photon energy.

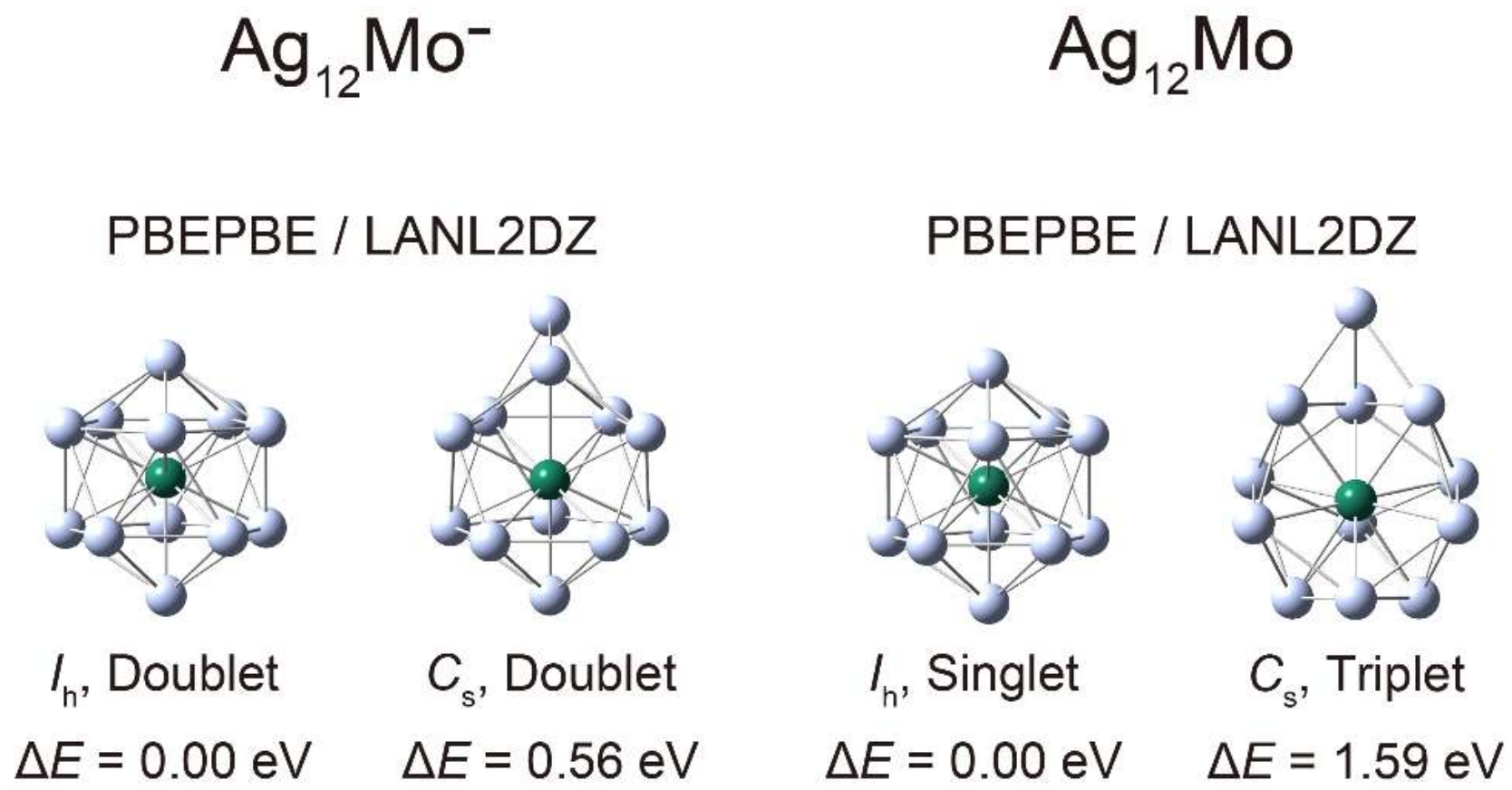

**Fig. S10. Optimized geometric structures of $Ag_{12}Mo^-$ and $Ag_{12}Mo$.**

The lowest and the second lowest energy isomers of anionic $Ag_{12}Mo^-$ (left columns) and neutral $Ag_{12}Mo$ (right columns) at PBEPBE/LANL2DZ level of theory. Their symmetry, spin state, and relative energy ($\Delta E$) are also presented below each structure.

**Table S1. Computational results of VDE and ADE for $Ag_{12}W^{-}$.**

VDE and ADE are listed for optimized geometries obtained by each computational level (in eV units). The same effective core potential and corresponding basis sets are used for Ag and W atoms.

| | | BP86 | PBEPBE | B3LYP | HSEH1PBE | PBE1PBE |
|---|---|---|---|---|---|---|
| LANL2DZ | VDE | 1.56 | **1.46** | 1.33 | 1.26 | 1.19 |
| | ADE | 1.56 | **1.46** | 1.33 | 1.26 | 1.19 |
| SDD | VDE | 1.70 | 1.59 | 1.43 | 1.36 | 1.29 |
| | ADE | 1.70 | 1.59 | 1.43 | 1.36 | 1.29 |
| cc-pVDZ-PP | VDE | 1.71 | 1.59 | 1.42 | 1.35 | 1.26 |
| | ADE | 1.71 | 1.59 | 1.42 | 1.35 | 1.26 |
| aug-cc-pVDZ-PP | VDE | 1.73 | 1.61 | 1.45 | 1.38 | 1.30 |
| | ADE | 1.72 | 1.61 | 1.45 | 1.38 | 1.30 |
| def2-SVPD | VDE | 1.66 | 1.55 | 1.40 | 1.31 | 1.24 |
| | ADE | 1.66 | 1.54 | 1.40 | 1.31 | 1.24 |
| def2-TZVPP | VDE | 1.68 | 1.56 | 1.40 | 1.32 | 1.24 |
| | ADE | 1.68 | 1.56 | 1.40 | 1.32 | 1.24 |
| def2-TZVPD | VDE | 1.69 | 1.57 | 1.41 | 1.33 | 1.25 |
| | ADE | 1.68 | 1.57 | 1.41 | 1.33 | 1.25 |
| **Experimental VDE** | | | | **1.46** | | |

**Table S2. Orbital angular momentum eigenstates vs. symmetry representations.** Irreducible representations of superatomic orbitals under the $I_h$ point group are listed in correspondence with orbital angular momentum eigenstates under full spherical symmetry. Here, $l'$ denotes the orbital angular momentum quantum number.

| Orbital angular momentum eigenstates under full spherical symmetry | Symmetry representations under $I_h$ point group |
|---|---|
| S ($l'$ = 0) | $A_g$ |
| P ($l'$ = 1) | $T_{1u}$ |
| D ($l'$ = 2) | $H_g$ |
| F ($l'$ = 3) | $T_{2u} \oplus G_u$ |
| G ($l'$ = 4) | $G_g \oplus H_g$ |
| H ($l'$ = 5) | $T_{1u} \oplus T_{2u} \oplus H_u$ |
| I ($l'$ = 6) | $A_g \oplus T_{1g} \oplus G_g \oplus H_g$ |

**References**

1. Frisch, M. J.; Trucks, G. W.; Schlegel, H. B.; Scuseria, G. E.; Robb, M. A.; Cheeseman, J. R.; Scalmani, G.; Barone, V.; Petersson, G. A.; Nakatsuji, H.; Li, X.; Caricato, M.; Marenich, A. V.; Bloino, J.; Janesko, B. G.; Gomperts, R.; Mennucci, B.; Hratchian, H. P.; Ortiz, J. V.; Izmaylov, A. F.; Sonnenberg, J. L.; Williams-Young, D.; Ding, F.; Lipparini, F.; Egidi, F.; Goings, J.; Peng, B.; Petrone, A.; Henderson, T.; Ranasinghe, D.; Zakrzewski, V. G.; Gao, J.; Rega, N.; Zheng, G.; Liang, W.; Hada, M.; Ehara, M.; Toyota, K.; Fukuda, R.; Hasegawa, J.; Ishida, M.; Nakajima, T.; Honda, Y.; Kitao, O.; Nakai, H.; Vreven, T.; Throssell, K.; Montgomery, J. A., Jr.; Peralta, J. E.; Ogliaro, F.; Bearpark, M. J.; Heyd, J. J.; Brothers, E. N.; Kudin, K. N.; Staroverov, V. N.; Keith, T. A.; Kobayashi, R.; Normand, J.; Raghavachari, K.; Rendell, A. P.; Burant, J. C.; Iyengar, S. S.; Tomasi, J.; Cossi, M.; Millam, J. M.; Klene, M.; Adamo, C.; Cammi, R.; Ochterski, J. W.; Martin, R. L.; Morokuma, K.; Farkas, O.; Foresman, J. B.; Fox, D. J. Gaussian16, Revision C. 01. (2016).
2. Becke, A. D. Density-functional exchange-energy approximation with correct asymptotic behavior. *Phys. Rev. A* **38**, 3098–3100 (1988).
3. Perdew, J. P. Density-functional approximation for the correlation energy of the inhomogeneous electron gas. *Phys. Rev. B* **33**, 8822–8824 (1986).
4. Perdew, J. P., Burke, K. & Ernzerhof, M. Generalized Gradient Approximation Made Simple. *Phys. Rev. Lett.* **77**, 3865–3868 (1996).
5. Adamo, C. & Barone, V. Toward reliable density functional methods without adjustable parameters: The PBE0 model. *J. Chem. Phys.* **110**, 6158–6170 (1999).

6. Krukau, A. V., Vydrov, O. A., Izmaylov, A. F. & Scuseria, G. E. Influence of the exchange screening parameter on the performance of screened hybrid functionals. *J. Chem. Phys.* **125**, 224106 (2006).

7. Hay, P. J. & Wadt, W. R. Ab initio effective core potentials for molecular calculations. Potentials for K to Au including the outermost core orbitals. *J. Chem. Phys.* **82**, 299–310 (1985).

8. Andrae, D., Häußermann, U., Dolg, M., Stoll, H. & Preuß, H. Energy-adjusted ab initio pseudopotentials for the second and third row transition elements. *Theoret. Chim. Acta* **77**, 123–141 (1990).

9. Martin, J. M. L. & Sundermann, A. Correlation consistent valence basis sets for use with the Stuttgart–Dresden–Bonn relativistic effective core potentials: The atoms Ga–Kr and In–Xe. *J. Chem. Phys.* **114**, 3408–3420 (2001).

10. Becke, A. D. Density-functional thermochemistry. III. The role of exact exchange. *J. Chem. Phys.* **98**, 5648–5652 (1993).

11. Lee, C., Yang, W. & Parr, R. G. Development of the Colle-Salvetti correlation-energy formula into a functional of the electron density. *Phys. Rev. B* **37**, 785–789 (1988).

12. Miehlich, B., Savin, A., Stoll, H. & Preuss, H. Results obtained with the correlation energy density functionals of becke and Lee, Yang and Parr. *Chem. Phys. Lett.* **157**, 200–206 (1989).

13. Weigend, F. & Ahlrichs, R. Balanced basis sets of split valence, triple zeta valence and quadruple zeta valence quality for H to Rn: Design and assessment of accuracy. *Phys. Chem. Chem. Phys.* **7**, 3297–3305 (2005).

14. Rappoport, D. & Furche, F. Property-optimized Gaussian basis sets for molecular response calculations. *J. Chem. Phys.* **133**, 134105 (2010).
15. Figgen, D., Rauhut, G., Dolg, M. & Stoll, H. Energy-consistent pseudopotentials for group 11 and 12 atoms: adjustment to multi-configuration Dirac–Hartree–Fock data. *Chem. Phys.* **311**, 227–244 (2005).
16. Figgen, D., Peterson, K. A., Dolg, M. & Stoll, H. Energy-consistent pseudopotentials and correlation consistent basis sets for the 5d elements Hf–Pt. *J. Chem. Phys.* **130**, 164108 (2009).
17. Peterson, K. A. & Puzzarini, C. Systematically convergent basis sets for transition metals. II. Pseudopotential-based correlation consistent basis sets for the group 11 (Cu, Ag, Au) and 12 (Zn, Cd, Hg) elements. *Theor. Chem. Acc.* **114**, 283–296 (2005).
18. Epifanovsky, E. *et al.* Software for the frontiers of quantum chemistry: An overview of developments in the Q-Chem 5 package. *J. Chem. Phys.* **155**, 084801 (2021).
19. Gozem, S. *et al.* Photoelectron Wave Function in Photoionization: Plane Wave or Coulomb Wave? *J. Phys. Chem. Lett.* **6**, 4532–4540 (2015).
20. Gozem, S. & Krylov, A. I. The *ezSpectra* suite: An easy-to-use toolkit for spectroscopy modeling. *WIREs Comput. Mol. Sci.* **12**, e1546 (2022).